%% file: main_Arxiv_Version.tex
\documentclass[sigconf, nonacm, language=english, anonymous=false]{acmart}
\usepackage[utf8]{inputenc}
\usepackage{url}
\usepackage{booktabs}
\usepackage{hyperref}

\usepackage{breakurl}
\usepackage{wrapfig}
\usepackage{graphicx}
\usepackage{subfig}
\usepackage{float}
\usepackage{multirow}
\usepackage{tikz}
\usepackage{pifont}
\newcommand{\xmark}{\ding{55}}

\begin{document}
\title{The Relationship Between Network Similarity and Transferability of Adversarial Attacks}

\author{Gerrit Klause}
\orcid{0009-0008-0702-8264}
\affiliation{%
	\institution{Fraunhofer SIT / ATHENE}
	\streetaddress{Rheinstraße 75}
	\city{Darmstadt}
	\state{Hesse}
	\country{Germany}
	\postcode{64295}
}
\email{klause@sit.fraunhofer.de}

\author{Niklas Bunzel}
\affiliation{%
	\institution{Fraunhofer SIT / ATHENE / TU-Darmstadt}
	\city{Darmstadt}
	\country{Germany}
}
\email{bunzel@sit.fraunhofer.de}
\orcid{0000-0002-8921-1562}

\begin{abstract}
	Neural networks are vulnerable to adversarial attacks, and several defenses have been proposed. Designing a robust network is a challenging task given the wide range of attacks that have been developed. Therefore, we aim to provide insight into the influence of network similarity on the success rate of transferred adversarial attacks. Network designers can then compare their new network with existing ones to estimate its vulnerability. To achieve this, we investigate the complex relationship between network similarity and the success rate of transferred adversarial attacks. We applied the Centered Kernel Alignment (CKA) network similarity score and used various methods to find a correlation between a large number of Convolutional Neural Networks (CNNs) and adversarial attacks. Network similarity was found to be moderate across different CNN architectures, with more complex models such as DenseNet showing lower similarity scores due to their architectural complexity. Layer similarity was highest for consistent, basic layers such as DataParallel, Dropout and Conv2d, while specialized layers showed greater variability. Adversarial attack success rates were generally consistent for non-transferred attacks, but varied significantly for some transferred attacks, with complex networks being more vulnerable. We found that a DecisionTreeRegressor can predict the success rate of transferred attacks for all black-box and Carlini \& Wagner attacks with an accuracy of over 90\%, suggesting that predictive models may be viable under certain conditions. However, the variability of results across different data subsets underscores the complexity of these relationships and suggests that further research is needed to generalize these findings across different attack scenarios and network architectures.
\end{abstract}


\maketitle

\section{Introduction}
\input{content/intro}
\section{Background}
\input{content/back}
\section{Related Work}
\input{content/sota}
\section{Evaluation}
\input{content/eval}
\section{Conclusion \& Future Work}
\input{content/conclusion}
\begin{acks}
	This research work has been funded by the German Federal Ministry of Education and Research and the Hessian Ministry of Higher Education, Research, Science and the Arts within their joint support of the National Research Center for Applied Cybersecurity ATHENE.
\end{acks}

\bibliographystyle{ACM-Reference-Format}
\bibliography{mybib}

\end{document}

%% file: content/intro.tex
In recent years, significant advancements have been made in machine learning and artificial intelligence. These advancements in artificial intelligence (AI) owe significantly to neural networks. These networks excel in tasks like image recognition and speech processing, finding applications in healthcare, consumer tech, and entertainment. In healthcare, neural networks assist in diagnosing diseases and predicting patient outcomes through medical image interpretation~\cite{medical_imaging}. In consumer technology, they power voice assistants such as Siri and Alexa, providing speech recognition~\cite{nassif2019speech} and response generation. Furthermore, companies like Netflix and Amazon employ neural networks for personalized content and product recommendations based on user preferences.

In the domain of image recognition and processing, the proficiency of neural networks has been remarkable in recent years. The improvements in this domain allow for real time object detection~\cite{LeeH22} or image classification~\cite{sharma2018analysis} and among others is used in surveillance systems~\cite{facial_recognition} and autonomous driving~\cite{RashmiC24, autonomous_driving, Bourzac2016}. Since these use cases raise public safety concerns, they require a high level of security. The danger of adversarial attacks is great because they can trick neural networks into making wrong predictions. Surveillance systems could be manipulated or autonomous cars could misinterpret traffic signs potentially leading to accidents. 

The emergence of adversarial attacks on neural networks marks a significant development, first noted in 2013~\cite{szegedyIntriguingPropertiesNeural2014}. These attacks, capable of misleading highly accurate models with minimal perturbations, often imperceptible by the human eye, have led to the development of various adversarial example generation methods~\cite{brownAdversarialPatch2018, moosavidezfooli2017universal, baundary_attack_bib, squareattack_bib, bunzel2023concise, pgd_bib}. The growing awareness of these attacks' potential for misuse has spurred a continuous effort to balance offensive and defensive strategies in this field.

Defensive strategies against adversarial attacks in neural networks are manifold~\cite{pgd_bib, XiaoZZ20, papernot2016distillation, bunzel2024signals, Xu0Q18}. The focus is on model robustness and reducing the impact of malicious perturbations. These strategies range from adversarial training~\cite{pgd_bib} to input pre-processing~\cite{das2017keeping, Xu0Q18, roth2019odds, bunzel2024adversarial}, regularization and ensemble methods~\cite{AbbasiG17, Xu0Q18, mani2019towards}. There is also a growing emphasis on comprehensive evaluation metrics and benchmarks to assess model robustness against adversarial threats~\cite{croce2020robustbench, dong2020benchmarking, li2021adversarial}. 

In parallel, metrics have also been developed to assess the similarity between networks~\cite{svcca_bib, cca_bib, cka_bib}. It can help to understand how networks make their predictions and find strengths and weaknesses of different models. The Centered Kernel Alignment similarity metric~\cite{cka_bib}, has recently become a popular approach and has been widely used to compare representations of different networks. 

Despite significant research, there is a gap in the metrics for evaluating the success of adversarial attacks, particularly concerning the intrinsic characteristics of targeted neural networks. This paper aims to address this gap by evaluating the influence of network similarity on the transferred attacks success rate. Prior research has explored network similarity, but the link between this similarity and vulnerability to adversarial attacks remains underexplored.
Hereby our main contributions are:

\begin{itemize}
    \item A comprehensive analysis of network and layer similarity across a wide range of torchvision models, utilizing the Centered Kernel Alignment (CKA) metric.
    \item A novel metric for assessing network similarity by focusing on the most comparable layers between networks.
    \item A thorough evaluation of attack success rates and transferred attack success rates using a variety of torchvision models, in conjunction with both white-box and black-box attacks from the Adversarial Robustness Toolbox (ART).
    \item A regression model capable of predicting transferred attack success rates under specific conditions.
\end{itemize}

%% file: content/back.tex
\subsection{Adversarial Attacks}
\label{sec:adversarial_attacks}

Adversarial attacks manipulate the neural network output by adding small, carefully crafted perturbations to input data with various objectives. These attacks target domains such as image classification~\cite{fgsm_bib}, object detection~\cite{yin2021adc}, natural language processing~\cite{zhang2019adversarial}, and speech recognition~\cite{zelasko2021adversarial}. Concerns in deep learning emerged when Szegedy et al.~\cite{fgsm_bib} discovered that minimal, often imperceptible, input perturbations could significantly alter the predictions of deep neural networks. This issue, first noted in image classification, is now recognized in other computer vision tasks like semantic segmentation~\cite{arnab2018robustness}, object detection~\cite{MI2023114}, and tracking~\cite{SUTTAPAK202221}. Adversarial perturbations are particularly concerning because they can mislead models into making incorrect predictions with high confidence and are effective across different models~\cite{transferability_0}. Adversarial perturbations are generated to misclassify a specific input. Universal perturbations are designed to be effective across multiple inputs and capable of fooling multiple models~\cite{moosavidezfooli2017universal}. They highlight significant security risks, especially given the high expectations from deep learning technologies. The field has seen extensive research over the past years.

Adversarial attacks are a significant threat to the security of machine learning models, especially in security-sensitive applications. The main risks of adversarial attacks include model failure, where these attacks cause deep learning models to malfunction~\cite{papernot2015limitations}. Theses attacks also raises ethical questions about the reliability and safety of machine learning models.

The attacks used for analysis are explained further in this section. 

To defend against evasion attacks, a range of methods have been developed, each addressing different aspects of neural network security. Adversarial training involves adding adversarial samples to the training set, thereby enhancing the model's robustness~\cite{fgsm_bib}. Network distillation leverages the concept of training specialized smaller networks from a larger one, improving the model's resistance to subtle disturbances~\cite{papernot2016distillation}. Adversarial example detection focuses on identifying potentially harmful inputs using specialized detection models before they reach the main model.
Input reconstruction methods transform input samples in various ways, like noising and denoising, to resist attacks while maintaining the model's classification functions~\cite{9010982}. Each of these methods contributes uniquely to strengthening AI systems against the threats posed by evasion attacks.

\paragraph{Whitebox Attacks}
\label{paragraph:white_box}
White box attacks are executed with a comprehensive understanding of the target model, including its architecture and all underlying parameters. In such scenarios, the attacker predominantly leverages the ability to precisely calculate the gradient of the loss function in relation to the input data~\cite{meta_adv_attacks}. Because of the attackers knowledge of the model's internal mechanisms, white box attacks generally surpass black box attacks in terms of effectiveness~\cite{bhambri2020survey}.

\paragraph{Black-Box Attack} 
\label{paragraph:black_box}
In black box attacks, the adversary operates without insight into the target model's internal structure or its specific parameters. This form of attack grants the aggressor access solely to the model's inputs and its corresponding outputs. During this process, the attacker systematically probes the network to understand the behavior of the model~\cite{meta_adv_attacks}. Another option is for attackers to use surrogate models to generate adversarial instances in the hope that the perturbations will transfer and be effective against the target model~\cite{Qin_Xiong_Yi_Hsieh_2023}.

\paragraph{Non-targeted Attack}
\label{paragraph:non_targeted}
Non-targeted adversarial attacks are designed without a specific classification objective. The primary goal is to lead a machine learning model to classify input data incorrectly into any category except the correct one. 

\paragraph{Targeted Attack}
\label{paragraph:targeted}
Targeted adversarial attacks focus on a predetermined classification outcome. These attacks require the model to incorrectly categorize input data into a chosen class~\cite{meta_adv_attacks}.

\subsubsection{Gradient-based Attacks}
\label{subsec:gradient_based_attacks}
Gradient-based attacks are a powerful subset of adversarial attacks that exploit the gradient descent mechanism central to neural network learning. These attacks adjust the input data based on the computed gradients of the loss function, leading the model to incorrect predictions with minimal changes.

\subsubsection{Optimization-based Attacks}
\label{subsec:optimization_based_attacks}
Similar to gradient-based attacks, Optimization-based attacks work by subtly manipulating the input data in a way that is imperceptible to humans, but enough to cause the model to make a mistake. Unlike gradient-based attacks, which rely on information about the model's gradients, optimization-based attacks use advanced optimization algorithms to search for the smallest possible perturbations that can still cause the model to misclassify the input~\cite{meta_adv_attacks}. 

\subsection{Similarity metrics}

Network similarity measures for neural networks are crucial in understanding how different networks learn and represent information. It can help to analyze the transferability of features, comparing of network architectures. These measures can broadly be categorized into two types: representational similarity and functional similarity. Both types offer unique insights into the inner workings of neural networks, but they focus on different aspects of network behavior~\cite{csiszárik2021similarity}. 

\paragraph{Representational Similarity}

Representation similarity measures focus on the comparison of activation patterns produced by neural networks as they process inputs. These measures aim to quantify how similarly two networks encode information. A popular approach to assessing representational similarity is Canonical Correlation Analysis (CCA)~\cite{cca_bib} and its variants such as Singular Vector Canonical Correlation Analysis (SVCCA)~\cite{svcca_bib} and Centered Kernel Alignment (CKA)~\cite{cka_bib}.

\paragraph{Functional Similarity}

Functional similarity takes a broader approach to comparing neural networks (NNs) than representational similarity, focusing on the similarity in their overall functionality rather than their specific representations. This perspective is crucial for addressing higher-level inquiries, such as the compatibility of one network's representations with another's~\cite{csiszárik2021similarity}. Research in functional similarity is not as developed as in the domain of representational similarity, where methods like Canonical Correlation Analysis (CCA) are more established.

\subsubsection{Canonical Correlation Analysis (CCA)}

CCA is a statistical method used to find the relationships between two sets of variables. In the context of neural networks, it has been adapted to find semantic embeddings of images by comparing the activation patterns of networks. The initial approach using CCA for neural networks aimed to understand how different layers of a network represent information and how these representations change across networks. By finding linear combinations of features in each set that are maximally correlated, CCA provides a measure of similarity that can reveal shared representations between networks, even if those networks have different architectures or were trained on different tasks ~\cite{cca_bib}.

\subsubsection{Singular Vector Canonical Correlation Analysis (SVCCA)}

Building on the foundation of CCA, SVCCA enhances the method by first reducing the dimensionality of the activation spaces using singular value decomposition (SVD). This reduction makes the analysis more robust to noise and irrelevant dimensions, focusing on the most significant modes of variation in the data. SVCCA has been particularly useful in comparing deep learning models, as it allows for a more efficient and interpretable comparison of high-dimensional activation patterns~\cite{svcca_bib}.

\subsubsection{Centered Kernel Alignment (CKA)}
\label{subsec:cka}
CKA is a technique that measures the similarity of representations in neural networks by comparing the similarity matrices of the activations induced by a set of inputs~\cite{cka_bib}. Unlike CCA and SVCCA, which rely on linear correlations, CKA can capture both linear and nonlinear relationships between representations. This feature makes CKA particularly powerful for comparing networks with complex, non-linear transformations. CKA has been used to assess similarity by evaluating how similarly networks respond to the same inputs, providing insights into the invariances learned by different models. Because of its strength in capturing the similarity of representations in neural networks, CKA is the chosen technique to compare the CNNs in this work. The following provides a more detailed explanation of how CKA works. 

%% file: content/sota.tex
The transferability of Adversarial Attacks is a significant part of research in the field of adversarial attacks and especially important considering the security of machine learning systems~\cite{transferability_0, transferability_1,transferability_2,transferability_3}. It refers to the ability of attackers to create adversarial examples for a specific artificial neural network and transfer these examples to different models with different architectures and training data while still achieving the same effect.
This phenomenon was first examined by Szegedy et al. in 2014~\cite{szegedyIntriguingPropertiesNeural2014} and followed up on by Goodfellow et al. in 2015~\cite{fgsm_bib} who suggested that the transferability is due to the adversarial perturbation being closely aligned with the model's weight vector. They used the MNIST and CIFAR-10 datasets but later in 2017 Liu et al.~\cite{transferability_0} show that this is not the case for models trained on ImageNet. They also extended the black box setting by not having information about the training and test sets. This information was available to other studies that explored the transferability by constructing substitute models for the black-box target model and explored other models such as decision tree, kNN, etc.~\cite{papernotPracticalBlackBoxAttacks2017,papernotTransferabilityMachineLearning2016}. Petrov et al.~\cite{transferability_1} showed that white-box attacks on similar model architectures also have similarities in their adversarial perturbations. This was confirmed in 2023 by Alvarez et al.~\cite{alvarezExploringTransferabilityAdversarial2023} who also recognized that achieving transferability for attacks in real-world scenarios is still a challenge. Research in this field is still ongoing and studies are also attempting to find more systematic and quantifiable ways to compare attack transferability, while also recognizing that more research is needed to find consistent, fair ways to compare transferability~\cite{transferability_1}.

%% file: content/eval.tex
In this chapter, we address three critical aspects of our study: network similarity, the success rate of transferred adversarial attacks, and the correlation between these two factors. First, we evaluate network similarity scores to gain an understanding of what contributes to network similarity. Next, we analyse the success rate of adversarial attacks when transferred between networks, providing insight into the robustness and vulnerability of these models. Finally, we examine the correlation between network similarity and the transfer success rate of adversarial attacks, to determine the extent to which network characteristics influence the transferability of attacks.

\subsection{Network similarity}
For our goal of finding a correlation between network similarity and adversarial attack success rate, we first need to analyze the network similarities to better understand what contributes to the similarity scores and why networks are similar.

When examining the network similarity scores with CKA, it becomes evident that the analyzed networks exhibit a moderate degree of similarity overall. Both the mean and median similarity scores are 0.45, suggesting that the central tendency of the scores is firmly moderate. The low standard deviation of 0.05 highlights a lack of significant variability among the similarity scores, meaning that most scores are tightly clustered around the mean. This consistency indicates that the networks do not diverge widely in their layer activations and therefore in their predictions. Additionally, the observed range, with a maximum similarity score of 0.57 and a minimum of 0.32, confirms that while there is some variation, it is relatively limited. These metrics collectively suggest that the networks analyzed are moderately similar to each other, with a stable and predictable degree of similarity across the board.

\begin{figure}
    \centering
    \includegraphics[width=0.99\linewidth]{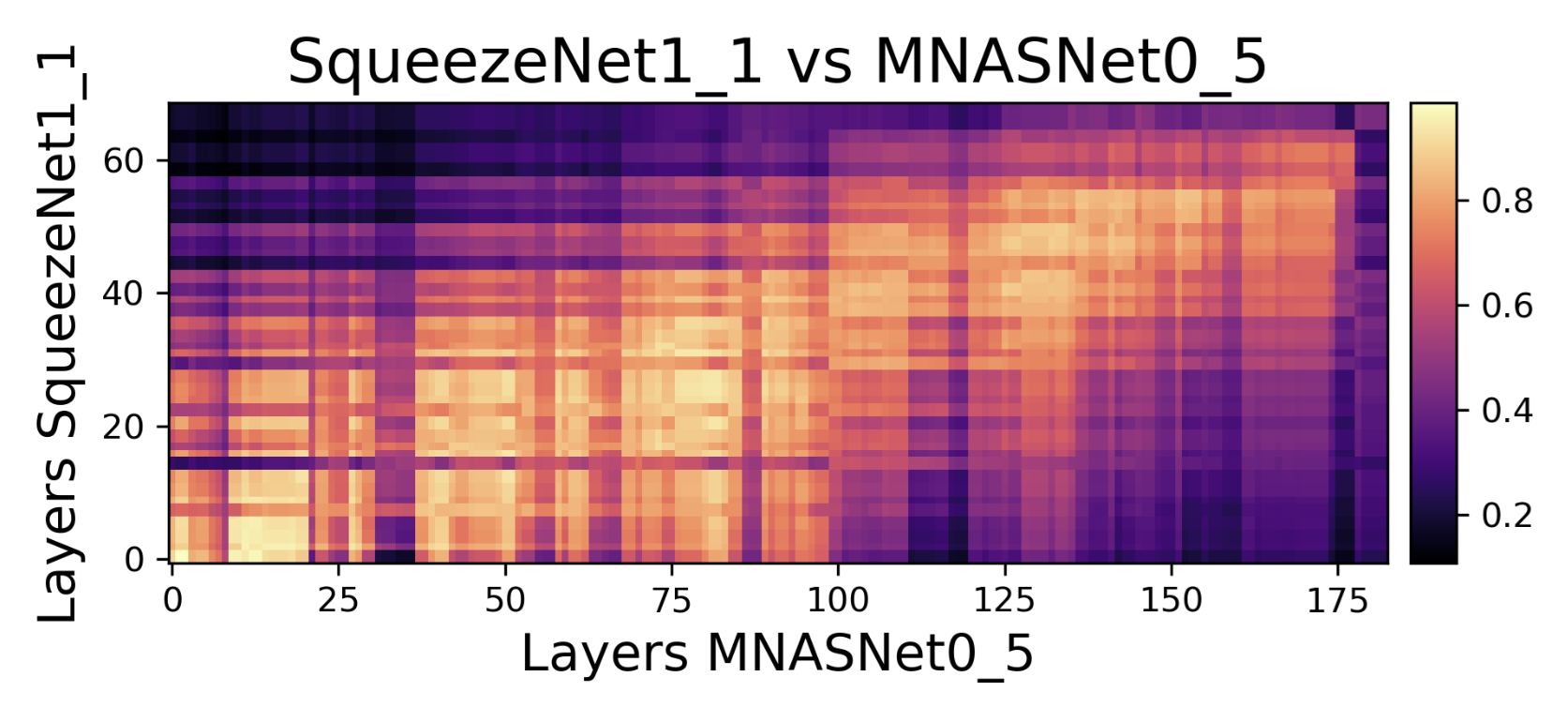}
    \caption{Similarity heatmap between SqueezeNet1\_1 and MNASNet0\_5. The similarity score is 0.57.}
    \label{fig:least_similar}
\end{figure}

\begin{figure}
    \centering
    \includegraphics[width=0.99\linewidth]{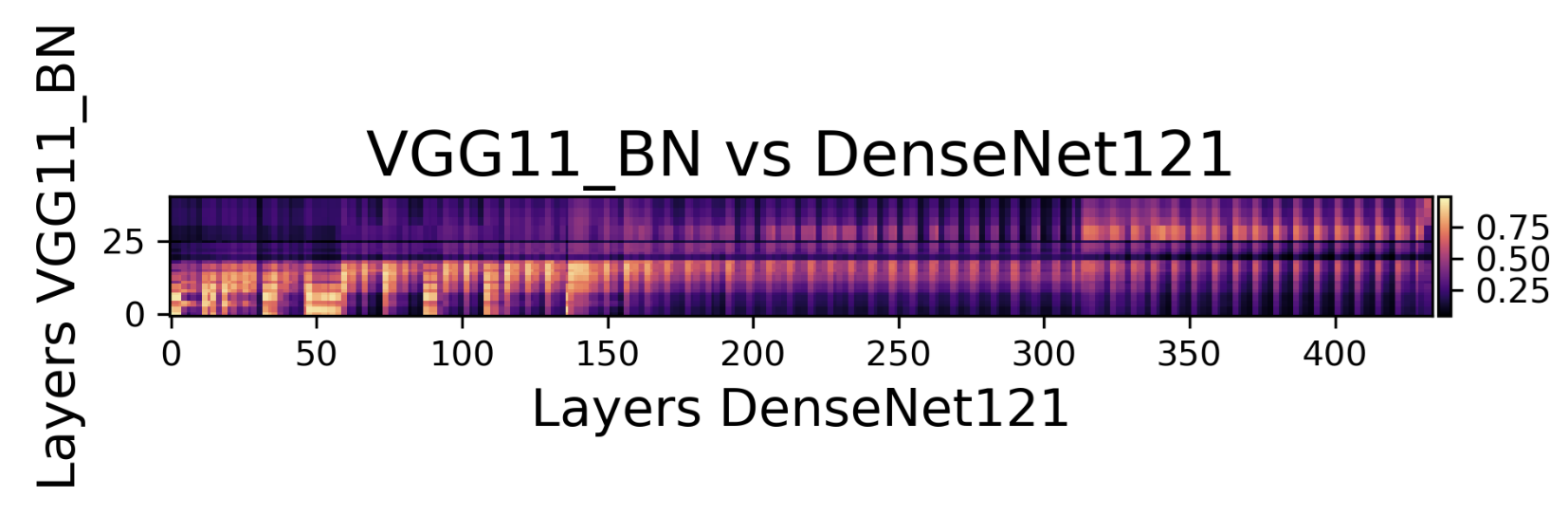}
    \caption{Similarity heatmap between VGG11\_BN and DenseNet201. The similarity score is 0.32.}
    \label{fig:most_similar}
\end{figure}

The most similar networks identified are SqueezeNet1\_1 and MNASNet0\_5, as highlighted in their similarity heatmap in~\ref{fig:most_similar}. This notable similarity can be attributed to their shared focus on efficiency, specifically targeting mobile and embedded device applications. SqueezeNet1\_1 employs a strategy of reducing the number of parameters through the use of 1x1 convolutions and deep compression without significantly compromising accuracy. Similarly, MNASNet0\_5 is designed with a neural architecture search framework that optimizes both accuracy and efficiency, balancing these goals by reducing computational cost and the number of parameters. Despite the use of different architectural components, their underlying goal of creating lightweight, high-performance networks for constrained environments likely leads to their observed similarity.

\begin{figure}
    \centering
    \includegraphics[width=0.99\linewidth]{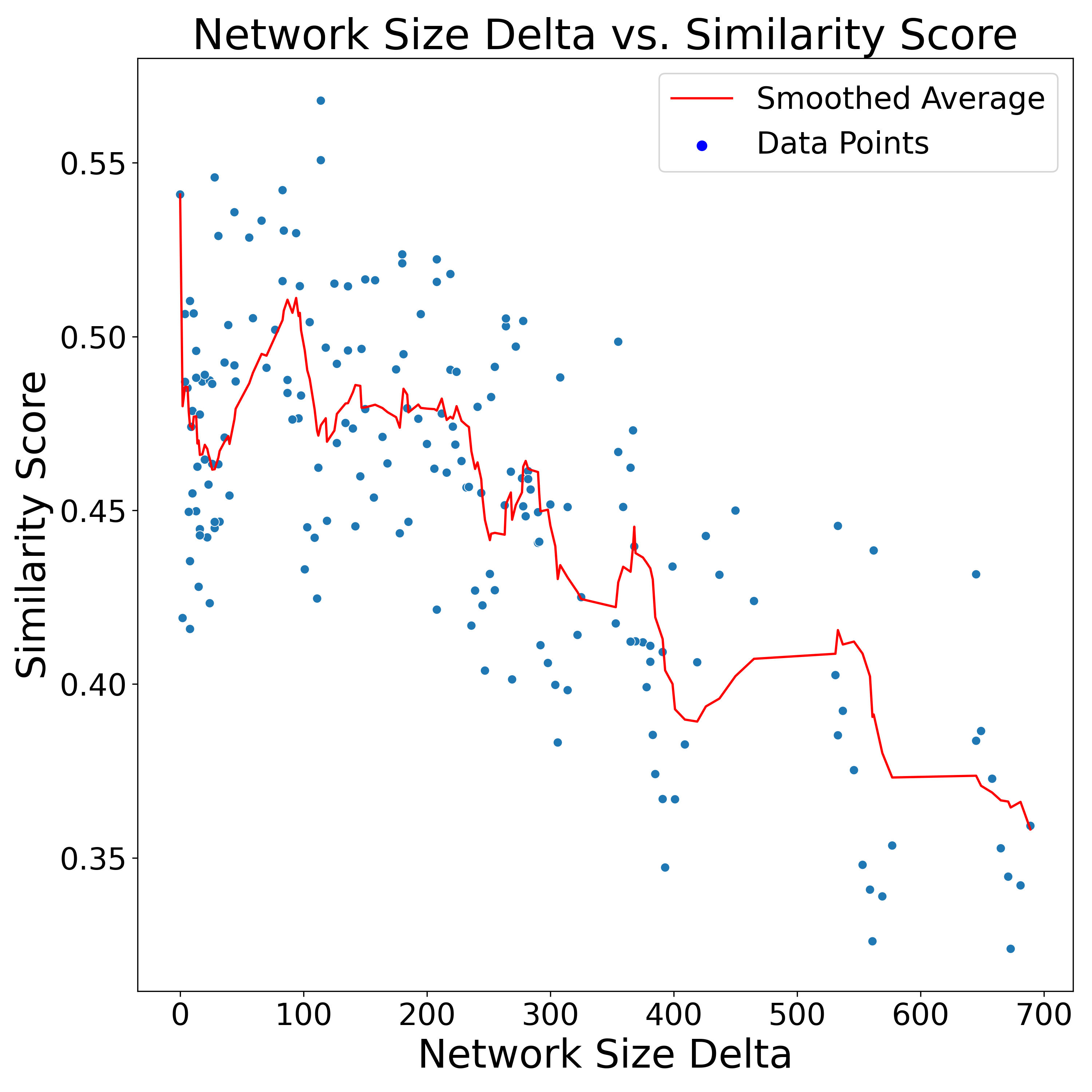}
    \caption{Difference in the number of layers between networks and their similarity scores.}
    \label{fig:size_delta}
\end{figure}

\begin{figure}
    \centering
    \includegraphics[width=0.99\linewidth]{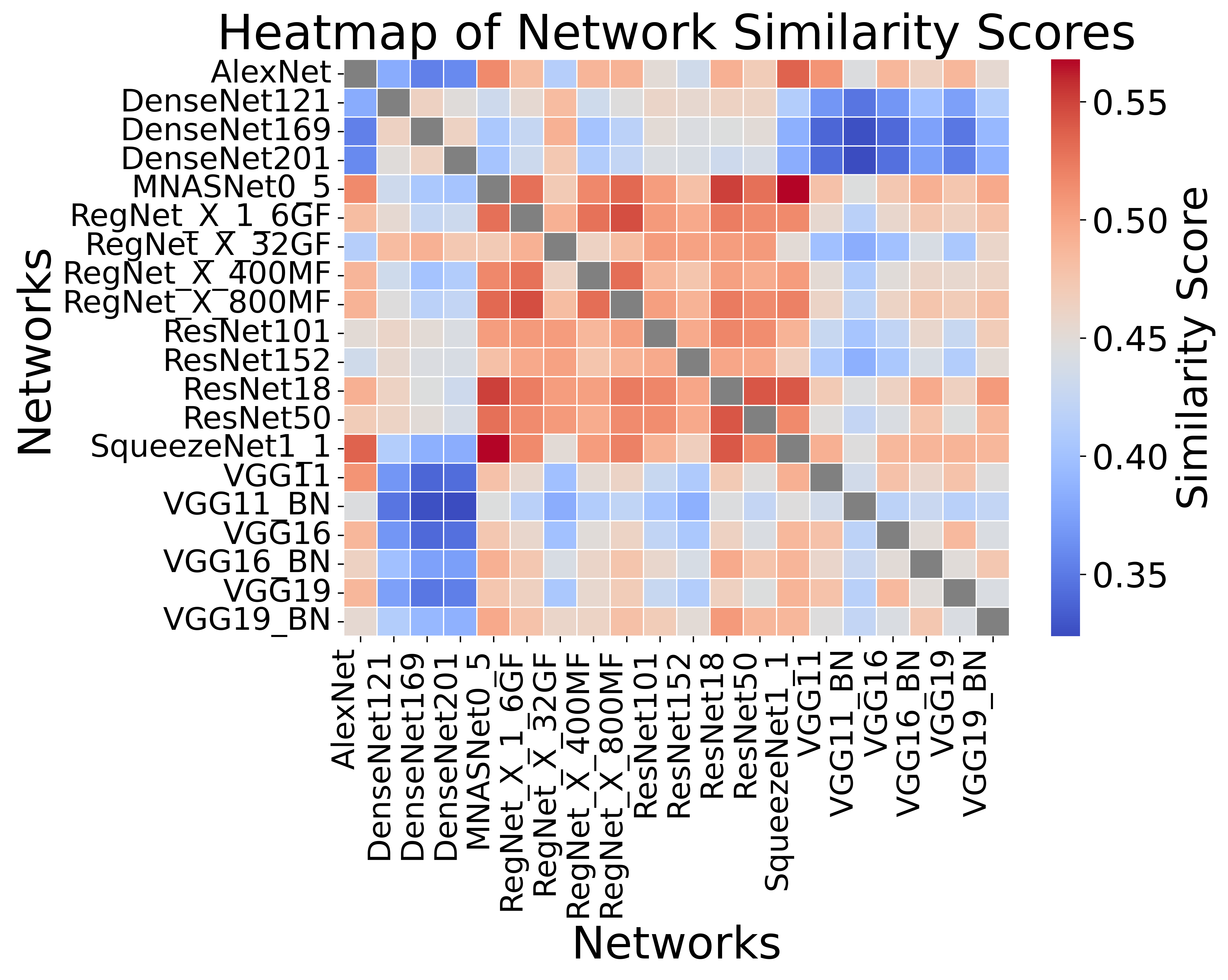}
    \caption{Heatmap of similarity scores between all pairs of networks.}
    \label{fig:all_similarities}
\end{figure}

The least similar networks identified are VGG11\_BN and \newline DenseNet201, as shown in their similarity heatmap in~\ref{fig:least_similar}. This disparity is likely due to their fundamentally different architectural designs and levels of complexity. VGG11\_BN is characterized by its simplicity and sequential architecture, making it one of the smaller models in terms of size. In contrast, DenseNet201 is a much deeper and more complex network with dense connections between layers, facilitating gradient flow and feature reuse. The significant difference in the number of layers and overall structure between these models contributes to their low similarity score. Analysis of the delta between the number of layers and the corresponding similarity scores, as shown in~\ref{fig:size_delta}, further supports the observation that similarity decreases as the size difference of networks increases. Furthermore, examination of the similarities across all networks in~\ref{fig:all_similarities} shows consistently low similarity scores for all DenseNet models compared to other architectures. This pattern suggests that the unique DenseBlock structure in DenseNet models inherently leads to lower similarity scores with other network types, emphasizing the distinctiveness of the DenseNet architecture.

When examining the similarity of each pair of networks, certain types of layers have remarkably low similarity scores. In particular, the VGG BN models contain one or more layers with similarity scores close to zero when compared to layers in other networks. The layers with low similarity scores are 2D convolutional layers, this unexpected result suggests that the low similarity scores are not due to the presence of BN layers. The exact reason why these particular 2D convolutional layers have such low similarity remains unclear, suggesting that there may be other factors or unique architectural features within the VGG BN models that influence these scores. Further investigation is required to uncover the underlying cause of this phenomenon.

We can also observe that the highest similarity scores are achieved at the beginning of the network in the early layers. This is likely because the early layers typically focus on extracting basic features such as edges and textures, which are common across different network architectures. As we move deeper into the network, the layers become more complex and specialized, dealing with higher level feature extraction and specific tasks, resulting in lower similarity scores. Interestingly, there is a noticeable increase in similarity scores in some of the final layers. This could be attributed to the fact that the final layers often converge towards similar structures, such as fully connected layers for classification purposes, which are more consistent across different network architectures. This pattern suggests that while the middle sections of networks diverge significantly in their complexity and specialization, both the initial and final layers tend to maintain a higher degree of similarity across different network designs.

Given these findings, it is crucial to further investigate the factors that contribute to the similarity scores. To achieve this, we will explore the different layer types and their similarity, as well as their respective positions within the networks. This approach will help us understand whether certain layer types consistently yield higher or lower similarity scores, and whether their position within the network affects their similarity to layers in other networks.

\subsection{Layer similarity}
\label{sec:layer_similarity}
To further explore the similarities and differences between neural networks, we now turn our attention to the individual layers within these networks. By comparing layers across networks, we can gain insight into the structural and functional components that contribute to network similarity or divergence. The analysis shows that the mean similarity score for layers across all networks is 0.45, indicating a moderate level of similarity on average. The median similarity score of 0.44, which is slightly lower than the mean, suggests that more than half of the layer comparisons have a similarity score below the mean, reflecting a skew in the distribution towards lower similarity.

The standard deviation of 0.20 is relatively high, indicating a wide range of similarity scores around the mean. This variability suggests that while some layers are quite similar, others are quite different. The maximum similarity score of 0.99 shows that there are pairs of layers that are almost identical, probably due to shared basic operations or common architectural components. Conversely, the minimum similarity score of 0.02 highlights the significant diversity in some layer comparisons where structural or functional differences are pronounced. Taken together, these statistics highlight the complex landscape of layer similarity in neural networks, driven by a mix of highly similar and highly divergent components.

\begin{figure}
    \centering
    \includegraphics[width=0.99\linewidth]{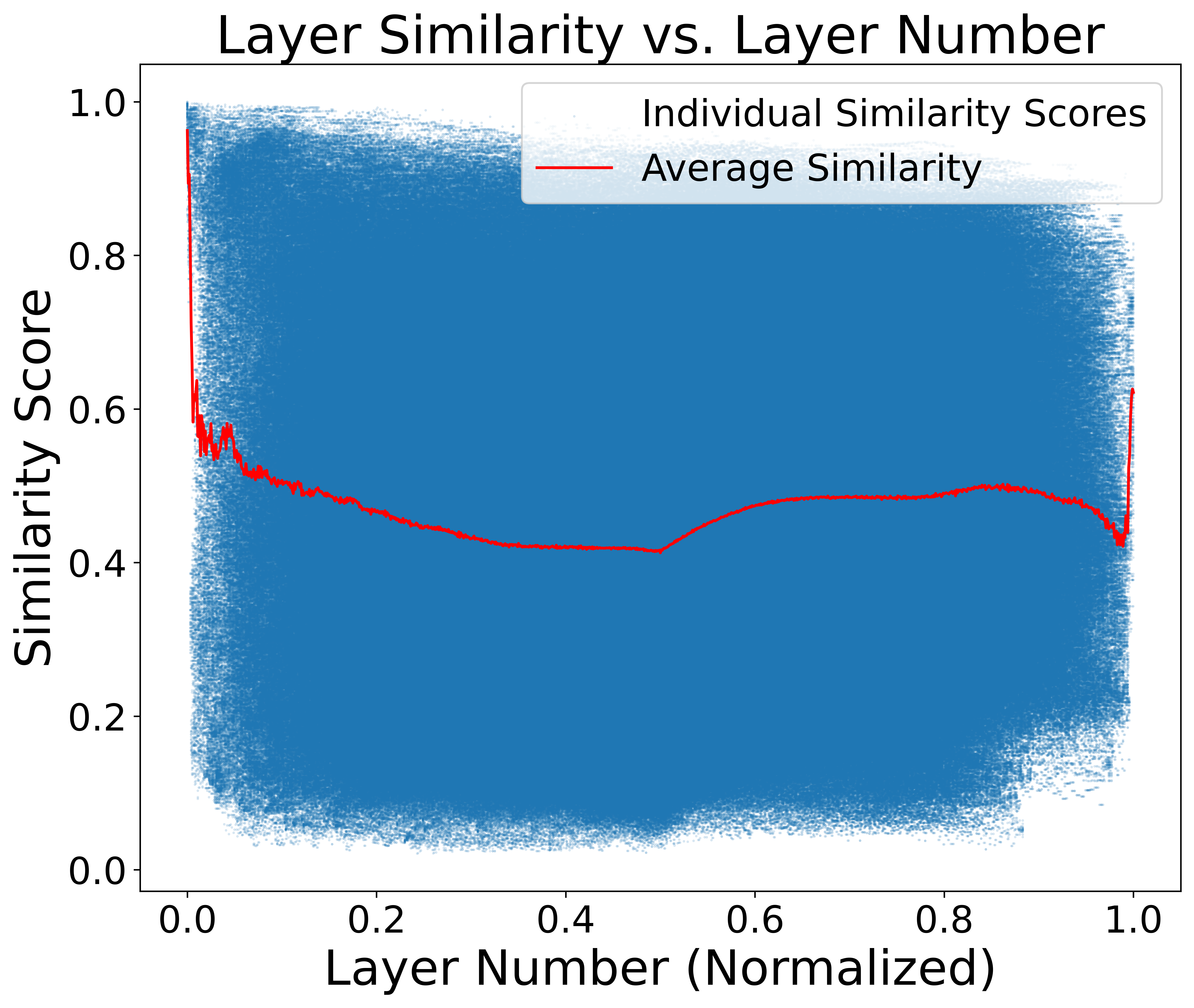}
    \caption{Mean similarity score for each layer position in the networks. The layer positions are normalized to 0-1, to allow for comparison across networks with different numbers of layers. The dot for each individual similarity score is not visible in the legend because the size is too small. This small size was chosen to make the figure more readable.}
    \label{fig:layer_position}
\end{figure}

Firstly, we will examine the average similarity scores based on the relative positions of each layer within each network, as illustrated in Figure~\ref{fig:layer_position}. The analysis reveals that the first layers consistently have the highest similarity scores, which aligns with the findings from the overall network similarity analysis. These initial layers typically focus on basic feature extraction, such as edges and textures, which are common across different networks and hence exhibit high similarity.
As we progress deeper into the network, the layers become increasingly complex and specialized to the specific tasks and architecture of the network, resulting in lower similarity scores. This complexity and specialization are reflected in the more varied and distinct layer functions and structures, leading to greater divergence.
Interestingly, the last layers show an increase in similarity scores. This can be attributed to the convergence of the networks towards similar final structures, such as fully connected layers used for classification purposes. These layers tend to exhibit greater uniformity across different networks, contributing to higher similarity scores towards the end of the network. This pattern highlights the structural and functional consistency of initial and final layers across diverse neural network architectures, with the intermediate layers showing greater variability and specialization.

\begin{figure}
    \centering
    \includegraphics[width=0.99\linewidth]{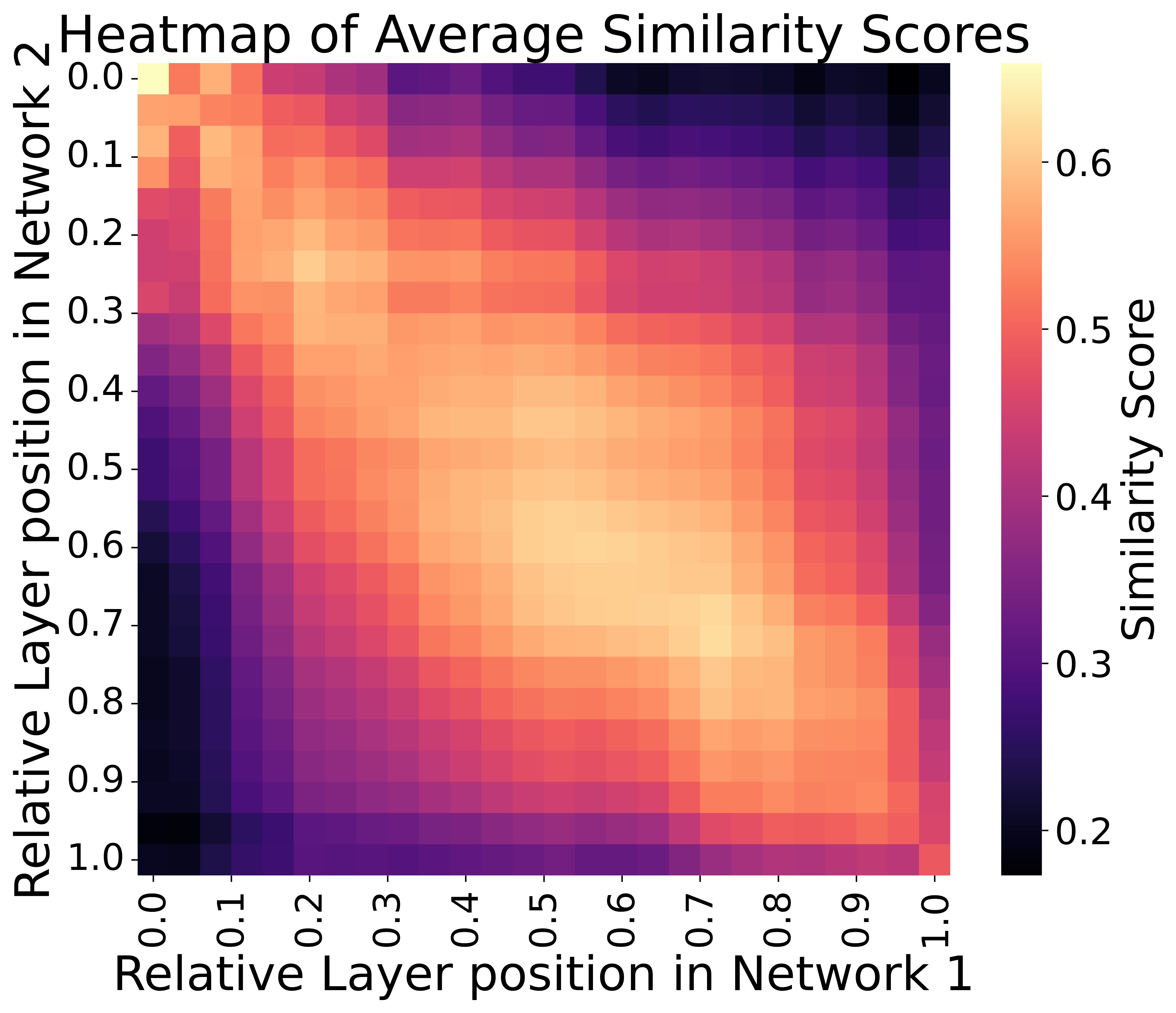}
    \caption{Heatmap of mean similarity scores between layers at there normalized positions in the networks.}
    \label{fig:heatmap_layer_position}
\end{figure}

Upon examination of the similarity scores for the normalized layer positions using Figure~\ref{fig:heatmap_layer_position}, it is again evident that the first layers exhibit the highest similarity scores. Notably, the highest similarity scores appear along the diagonal of the similarity matrix, indicating that layers positioned similarly within their respective networks tend to have higher similarity scores. This pattern is expected, as these layers are likely to perform similar functions and possess similar structures.
As we move further from the diagonal, the similarity scores decrease, indicating that layers positioned further apart in the networks are less similar to each other. In this visualisation of the similarity scores, the increase in similarity towards the final layers is less pronounced than in the previous analysis. This can be attributed to the grouping of more layers in the final cell, which does not contradict the previous results.

\begin{figure}
    \centering
    \includegraphics[width=0.99\linewidth]{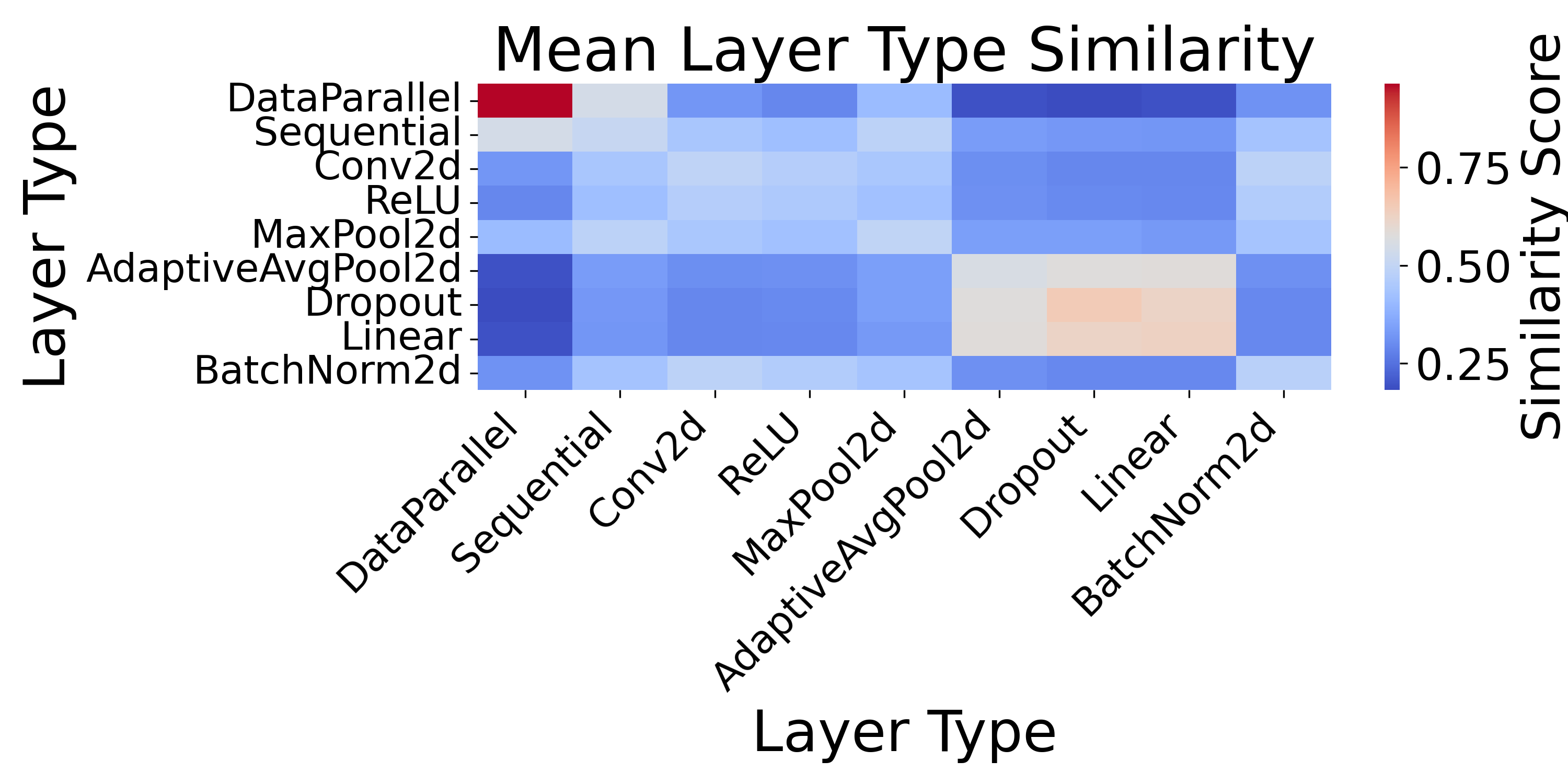}
    \caption{Heatmap of mean similarity scores between different types of layers. Only includes layer types that are present in more than one network architecture.}
    \label{fig:layer_types}
\end{figure}

The similarity scores for different types of layers will now be examined. The results are shown in Figure~\ref{fig:layer_types}.
The most obvious result is that DataParallel layers have the highest similarity scores. This is expected since they consistently appear as the first layer in each network, exclusively occupying that position and performing the same function across all networks. 
The next most similar layers are the Dropout, Linear and AdaptiveAvgPool2d layers. These layers are frequently used for regularization, classification, and feature extraction, respectively. Their higher similarity scores can be attributed to their similar functional roles across different networks. These layers typically appear at the same positions in the networks, mostly towards the end, which contributes to their high similarity scores. This positional consistency helps explain the high similarity scores observed in the first layers and the increase in similarity scores towards the end of the networks.
Furthermore, the Conv2d, ReLU, and BatchNorm2d layers also exhibit relatively high similarity. This is likely due to their ubiquitous presence in convolutional neural networks. These layers are foundational components in the architecture of CNNs and are frequently used in combination with each other, leading to higher similarity scores. Their common usage and fundamental role in building CNNs contribute to the observed pattern of similarity.
The analysis indicates that the highest similarity scores are associated with layers performing basic, common functions that are positioned consistently across networks. In contrast, more specialized and intermediate layers show greater variability and lower similarity scores. This helps to explain the observed trends in layer similarity across different network architectures.

\subsection{Diagonal Box Similarity}
Given these findings, it would be advantageous to extend the analysis by incorporating a network similarity score that solely considers layers situated in proximity to the diagonal of the similarity matrix. This refined metric will focus on the most comparable layers across networks, thereby providing a clearer understanding of their core similarities and differences while ignoring weakly comparable layers. By identifying the layers that contribute most significantly to network similarity, this approach can facilitate a deeper understanding of the structural and functional distinctions between different architectures. Furthermore, this refined analysis may assist in identifying correlations between network similarity and the success rates of adversarial attacks.

We will call this proposed method the \textit{Diagonal Box Similarity} (DBS) score. The DBS score will be calculated by using the \textit{Bresenham Line Algorithm}~\cite{bresenham} to get the points on the diagonal of the similarity matrix. Then considering a box of a certain size around each point on the diagonal and taking the mean of the similarity scores of all these points. 
Let $S \in \mathbb{R}^{n \times m} $ be the similarity matrix where $S_{ij}$ is the similarity score between layer $i$ of the first neural network and layer $j$ of the second neural network with $n,m$ being the amount of layers in the networks.
Let $(i_k, j_k) \in Bresenham \subseteq S$
be the points on the diagonal of the similarity matrix and lets use an empty set to keep track of unique points, $UniquePoints = \emptyset$.
Now for each point $(i_k, j_k) \in Bresenham$ we define the box
\[
    B = \left\{ 
    \begin{aligned}
        (i, j) \in S \mid & \max(1, i_k - r) \leq i \leq \min(m, i_k + r), \\
                          & \max(1, j_k - r) \leq j \leq \min(n, j_k + r)
    \end{aligned}
    \right\}
\]
and each point $(i,j) \in B$ gets added to UniquePoints, $UniquePoints \cup= \{(i, j)\}$.
Accumulate the absolute CKA similarity scores for the unique points,
\[
	\text{Sim} = \sum_{(i, j) \in \text{UniquePoints}} | S_{i,j} |
    \]
and divide by the number of unique points, \[\text{DBS}= \frac{\text{Sim}}{|\text{UniquePoints}|}\].

The DBS score 
\[
    \text{DBS} = \frac{\sum_{(i, j) \in \bigcup_{(i_k, j_k) \in \text{Bresenham}} B(i_k, j_k)} | S_{i,j} |}{\left\lvert \bigcup_{(i_k, j_k) \in \text{Bresenham}} B(i_k, j_k) \right\rvert}
	\]
will be a value between 0 and 1, where 0 indicates no similarity and 1 indicates perfect similarity. 

\begin{figure}
    \centering
    \includegraphics[width=0.99\linewidth]{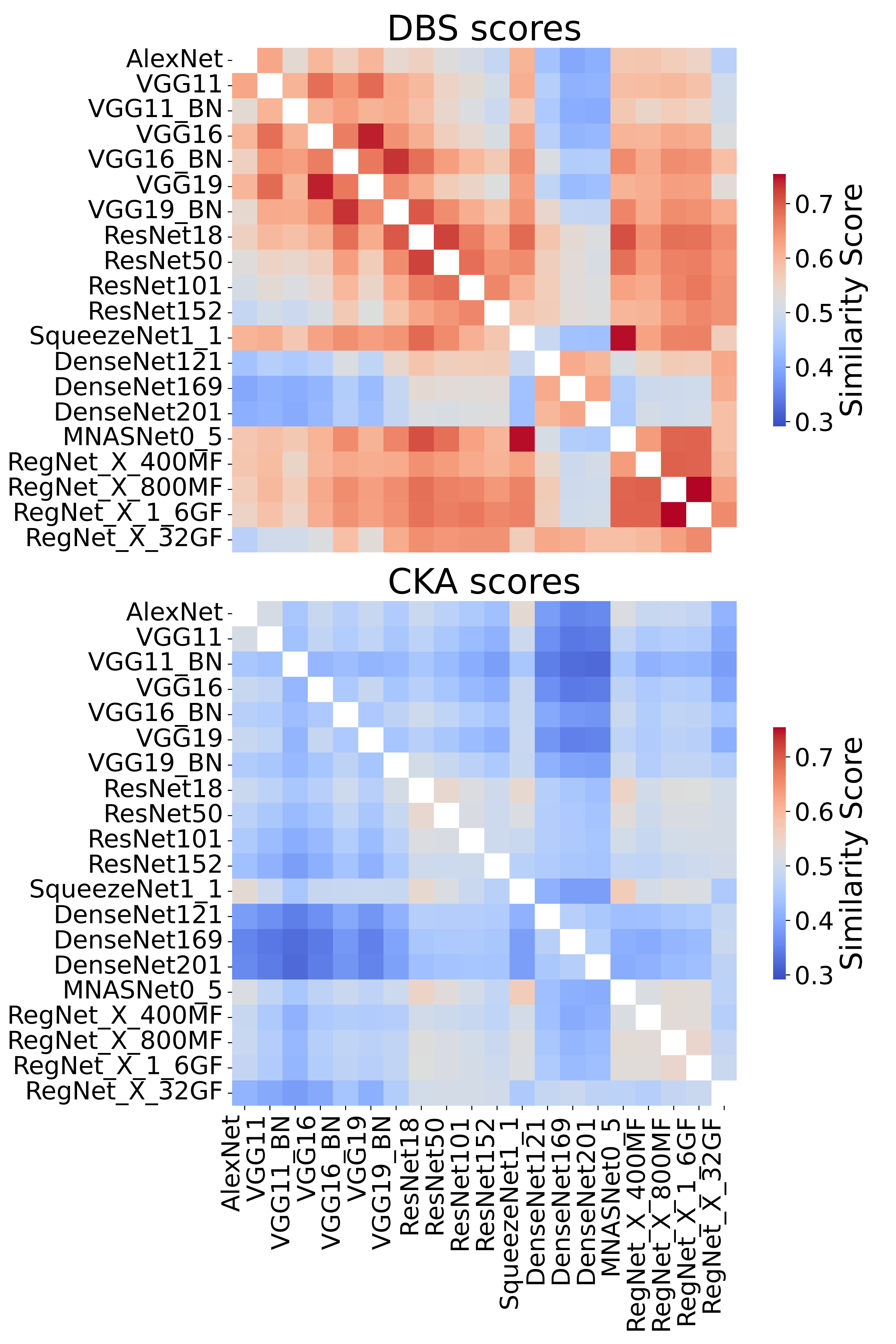}
    \caption{Similarity heatmaps for DBS and CKA scores with a vmin of 0.32 and a vmax of 0.75.}
    \label{fig:sim_comparison}
\end{figure}

We create the similarity matrix for all networks using the DBS score with different box sizes (1, 2, 3, 5, 7, 9, 11, 13, 15, 30, 45, 60, 75, 90, 105, 120, 150, 200, 300). The larger the box size chosen, the narrower the range of DBS scores. Looking at the heatmaps for each box size, we can see that a size of 5 seems to be a good middle ground between capturing the most important similarities and having a wider range of similarity scores to differentiate between networks. At this size, the same architectures such as VGG, ResNet and RegNet show high similarity. The heatmap for a box size of 5 in comparison to the CKA similarity is shown in Figure~\ref{fig:sim_comparison}. The DBS scores with box size 5 range from 0.40 to 0.75, while the CKA scores range from 0.32 to 0.57. We can see that the DBS heatmap shows a more varied color distribution with stronger red hues in VGG and ResNet areas. The CKA heatmap is predominantly blue because the value range is smaller.

For all further correlation analysis we will look at both the CKA and DBS scores to see if the DBS score can provide additional insight into the similarities and differences between networks.

\subsection{Adversarial attacks}
We will now look at the success rates of the adversarial attacks on the different networks. This analysis will help us to understand the different degrees of vulnerability of different convolutional neural network (CNN) architectures.

When we analyze the non-transferred attacks and their success rates on the individual networks, we can see that they differ slightly. 
Interestingly, if we take the mean of the success rates for each non-transferred attack across all networks, they are almost identical with $\sim$ 0.87 and a standard deviation of $\sim$ 0.04. This small standard deviation shows how little the success rates differ across networks, see~\ref{tab:mean_attacks}.
For most attacks, the success rates are very similar across all networks. 
This is true for both targeted and non-targeted attacks.
An additional correlation analysis with a pearson coefficient of $\sim$ 1 further supports this.

\begin{table}[h!]
    \centering
	\begin{tabular}{lcc}
	\textbf{Attacks} & \textbf{Mean Success} & \textbf{Mean Std} \\
	\hline
	Targeted Attacks & 0.870598 & 0.04 \\
	\hline
	Non-Targeted Attacks & 0.870656 & 0.04 \\
	\hline
	Overall & 0.870627 & 0.04 \\
	\hline
    \end{tabular}
    \caption{Comparison of non-transferred Attacks success rates on all Networks.}
    \label{tab:mean_attacks}
\end{table}

The first significant difference in success rate can be seen when looking at transferred attacks, see table~\ref{tab:adv_attacks_success_rates}. 
For non-targeted transferred attacks, PGD has the largest difference in success rate between networks, with the lowest success rate on Alexnet 0.75 and the highest on RegNet\_X\_32GF 0.92. PGD, FGSM and AutoCG are the only attacks that have a mean success rate lower than 90\% when transferred to all networks. All other attacks show very high transferability with a mean success rates of 90\%. The highest deviations can be seen for non-targeted FGSM with 22\% and PGD with 19\% non-targeted and 12\% targeted. All other attacks, even targeted FGSM, show a standard deviation of less than 7\%, indicating that they have very similiar transferability over all networks. 

\begin{table}[h!]
    \centering
    \begin{tabular}{lccc}
	\hline
	\textbf{Attack Name}             & \textbf{Targeted} & \textbf{Mean SR}   & \textbf{STD} \\ \hline
	Projected Gradient Descent  & \xmark            & 46\%          & 19\% \\
	Projected Gradient Descent  & \checkmark        & 75\%           & 12\%  \\
	Fast Gradient Sign Method       & \xmark            & 64\%           & 22\% \\
	Fast Gradient Sign Method       & \checkmark        & 83\%           & 7\%  \\
	Auto Conjugate Gradient          & \xmark            & 89\%           & 6\%  \\
	Auto Conjugate Gradient          & \checkmark        & 89\%            & 5\% \\
	Carlini L0 Method                & \xmark            & 92\%           & 3\% \\
	Carlini L0 Method                & \checkmark        & 93\%           & 3\%  \\ 
	Carlini L2 Method                & \xmark            & 93\%           & 3\% \\
	Carlini L2 Method                & \checkmark        & 93\%           & 3\%  \\
	DeepFool Attack                  & \xmark            & 92\%           & 3\% \\
	Spatial Transformation           & \xmark            & 94\%            & 3\% \\
	Square Attack                    & \checkmark        & 94\%            & 3\% \\
	Boundary Attack                  & \checkmark        & 94\%            & 3\% \\ \hline
    \end{tabular}
    \caption{Mean Success Rates and Standard Deviation of transferred Adversarial Attacks on all Networks.}
    \label{tab:adv_attacks_success_rates}
\end{table}

\begin{figure}
    \centering
    \includegraphics[width=0.99\linewidth]{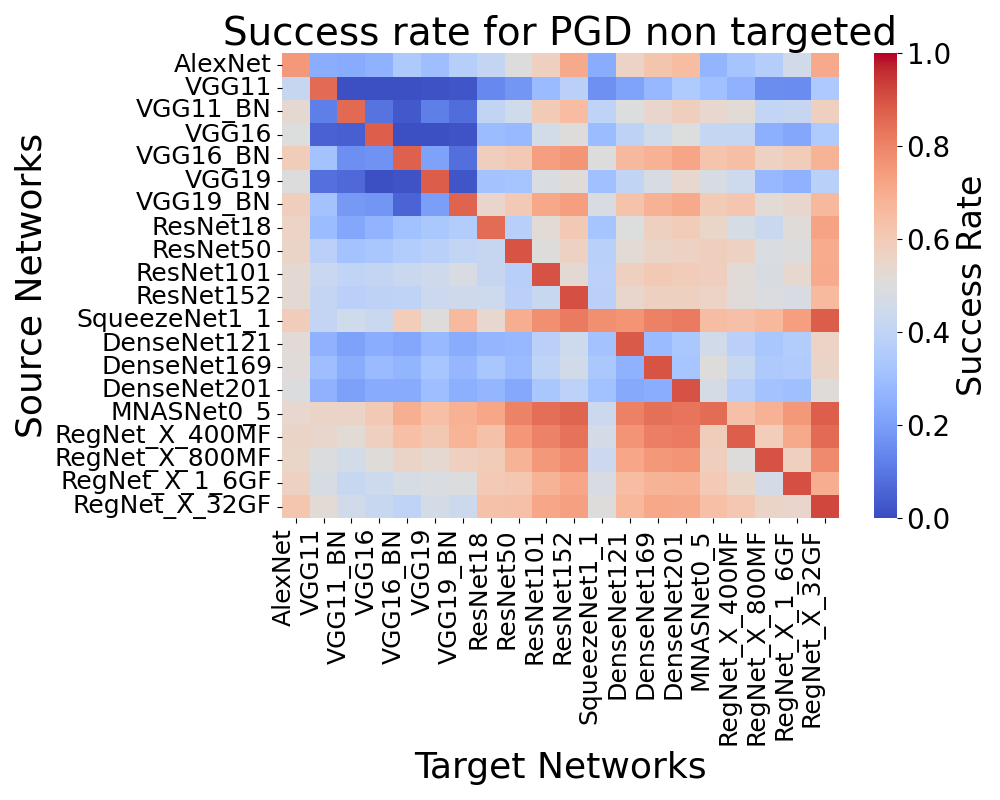}
    \caption{Success rate for non-targeted PGD attack on all networks.}
    \label{fig:fgsm_untargeted}
\end{figure}

\begin{figure}
    \centering
    \includegraphics[width=0.99\linewidth]{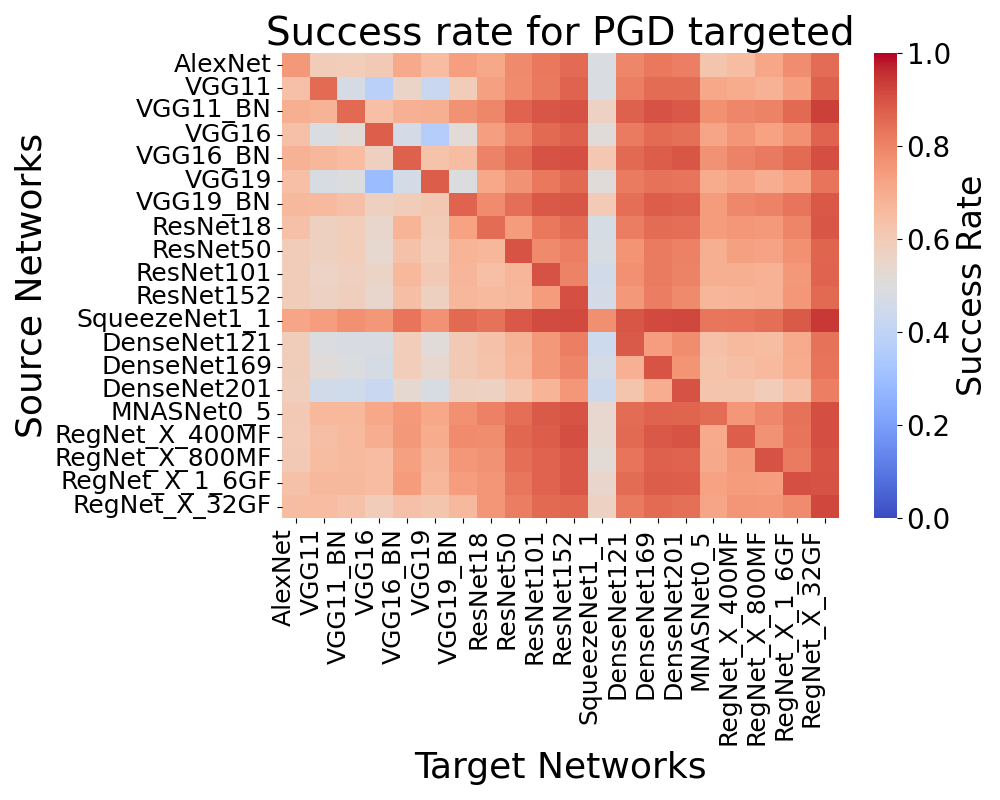}
    \caption{Success rate for targeted PGD attack on all networks.}
    \label{fig:fgsm_targeted}
\end{figure}

The biggest difference between targeted and non-targeted can be seen with transferred PGD, see figure~\ref{fig:fgsm_untargeted} and~\ref{fig:fgsm_targeted}. We can see that the success rates decrease significantly when the attack is non-targeted. This is true for all networks. It can be seen that the lower the success rate for the targeted attack, the greater the difference to the even lower success rates for the non-targeted attack.  
This is also reflected in the correlation analysis with a pearson coefficient of 0.39, indicating a small linear correlation between the success rates of the targeted and non-targeted transferred attacks. 

	
When looking at the transferability to certain networks, we see that RegNet\_X\_32GF is the most transferable to for all 14 out of 14 attacks. The least transferable to is AlexNet for 8 out of 14 attacks. 
This observation suggests that the architectural design of RegNet\_X\_32GF makes it more vulnerable to transferred attacks. The least transferable to are AlexNet, VGG11, and SqueezeNet1\_1 suggesting that simpler networks are more robust against transferred attacks from other networks.

When looking at the transferability from certain networks, we see that AlexNet is the most transferable from for 8 out of 14 attacks. The least transferable from network is RegNet\_X\_32GF for 8 out of 14 attacks. 
This observation suggests that attacks generated on simpler networks like AlexNet or small networks like SqeezeNet1\_1 are more transferable to other networks. The least transferable attacks are generated on more complex networks like RegNet\_X\_32GF or DenseNet201. 

So when creating attacks on a network, it is more likely that they are transferable to other networks if the network is simpler or smaller. When looking at the transferability to a network, it is more likely that the attack is successful if the network is more complex or larger. This aligns with recent studies about the transferability of evasion attacks~\cite{ambra}, the only outlier in our evaluation is VGG11 which is the least transferable from for targeted PGD and FGSM even though it is considered a rather small network.

Taking a closer look at the network mean success rate and standard deviation for transferred attacks, summarized in Table \ref{tab:cnn_transf_stats} we can confirm the findings. Notably, the RegNet\_X\_32GF network shows the highest mean success rate of 87.26\% with a standard deviation of 8.11\%, indicating a consistently high vulnerability to transferred attacks. In contrast, SqueezeNet1\_1 displayed the lowest mean success rate of 68.02\% and a standard deviation of 13.88\%, suggesting greater robustness against these attacks. The larger, more complex models also appear to be consistent across models in their high vulnerability to transferred attacks, as indicated by their low standard deviation. Smaller models like the VGG group show a higher standard deviation, indicating a higher variability in transferred attack success rates.

\begin{table}[h!]
    \centering
    \begin{tabular}{lcc}
	\hline
	\textbf{Network} & \textbf{Mean SR} & \textbf{STD} \\ \hline
	AlexNet & 70.94\% & 8.39\% \\ 
	VGG11 & 76.12\% & 16.57\% \\ 
	VGG11\_BN & 75.49\% & 17.60\% \\ 
	VGG16 & 76.39\% & 16.40\% \\ 
	VGG16\_BN & 77.44\% & 16.47\% \\ 
	VGG19 & 77.61\% & 16.70\% \\ 
	VGG19\_BN & 78.46\% & 15.36\% \\ 
	ResNet18 & 79.08\% & 12.34\% \\ 
	ResNet50 & 83.33\% & 12.51\% \\ 
	ResNet101 & 85.87\% & 9.98\% \\ 
	ResNet152 & 86.70\% & 8.59\% \\ 
	SqueezeNet1\_1 & 68.02\% & 13.88\% \\ 
	DenseNet121 & 82.74\% & 10.34\% \\ 
	DenseNet169 & 85.03\% & 9.32\% \\ 
	DenseNet201 & 85.18\% & 8.85\% \\ 
	MNASNet0\_5 & 78.44\% & 11.69\% \\ 
	RegNet\_X\_400MF & 80.15\% & 11.88\% \\ 
	RegNet\_X\_800MF & 80.71\% & 13.02\% \\ 
	RegNet\_X\_1\_6GF & 83.26\% & 12.53\% \\ 
	RegNet\_X\_32GF & 87.26\% & 8.11\% \\ \hline
    \end{tabular}
    \caption{Mean Success Rate and Standard Deviation for Each Network when Attacks are transferred to it.}
    \label{tab:cnn_transf_stats}
\end{table}

Considering these findings it might prove difficult to find a correlation between network similarity and the success of adversarial attacks. The success rates are very similar across all networks, with only minor differences. The most significant difference can be seen when looking at the transferability of the attacks, which is the parameter we try to predict.

\subsection{Correlation between network similarity and adversarial attacks}
\label{sec:correlation}
In this section, we will discuss in detail all the methods used to find a correlation between network similarity and the success rates of the adversarial attacks presented.
We used various correlation and analysis methods:
\begin{itemize}
    \item \textbf{Linear correlation methods:} Pearson's correlation.
    \item \textbf{Rank correlation methods:} Spearman's rank correlation, Kendall's tau.
    \item \textbf{Non-linear correlation methods:} Distance correlation.
    \item \textbf{Supervised learning:} Decision trees.
    \item \textbf{Visual correlation:} Scatter plots, heat maps.
\end{itemize}

To account for all possible correlations, we applied each of these methods to the different data subsets. We created subsets for different types of attacks, such as targeted and non-targeted attacks, black-box and white-box attacks, and single-step and multi-step attacks. We also created two subsets for models that have fewer than 200 layers and more than 200 layers. 

When looking at the correlation, we are trying to find a relationship between the similarity scores and the success rates of the submitted adversarial attacks, trying to find a measure of how well the similarity scores can predict the success rates of the submitted adversarial attacks. 
Another parameter that could be considered is the initial success rate of the attacks on the source model, but we did not consider this parameter in our analysis as it is a parameter that is most likely not available in a real world scenario.

We ran all these methods on the CKA and DBS similarity scores and found no significant difference.
Splitting the data into the subsets of small and large networks did not prove useful. There was no significant difference for all methods used.

\subsubsection{Correlation Methods}

Our first step to explore correlation, was computing several types of correlation coefficients: Pearson, Spearman, Kendall, and Distance Correlation all using the \texttt{SciPy} library~\cite{scypi} implementations.
The Pearson correlation coefficient measures the linear relationship between two continuous variables.
The Spearman rank correlation coefficient assesses the monotonic relationship between two variables. It is based on the ranks of the data rather than the raw data values, and was calculated using the \texttt{spearmanr} function.
The Kendall rank correlation coefficient measures the ordinal association between two variables.
Distance Correlation evaluates both linear and non-linear associations between two variables.
We used these to see if there was a general trend between the similarity scores and the success rates of the transferred adversarial attacks. These methods were used as indicators to highlight certain areas that might be worthy of further investigation, we used the absolute values of these methods with 0 representing no correlation and 1 indicating a strong correlation. This approach also accounts for anti-proportional correlations.
We took each network as source network and compared the similarity scores with the success rates of the transferred adversarial attacks for every other networks. We then calculated the correlation scores for each attack and each network. 
All these methods produced similar results. 
The results show a possible correlation for the smaller networks such as AlexNet and the VGG models, as well as for the DenseNet models. The ResNet and RegNet models show no correlation with these methods. Although the results indicate a relationship, it is only a general trend that deserves further investigation. The results could also be insignificant due to the small difference in the success rates of the attacks. A notable finding is that the results indicate that there does not appear to be a single attack that has a consistent correlation across all models. Rather, it is a particular model architecture that correlates with the success rates of the different transferred adversarial attacks. The general trends are the same for each method, just the intensity of the correlation varies slightly.

Looking at the difference between the CKA similarity scores and our DBS scores, we found that the trends are similar, but the DBS scores show a higher fluctuation in the correlation scores. The highest correlation values using CKA similarity scores were even higher using DBS scores, although on average the correlation values using DBS scores were lower. This suggests that these correlation methods are more sensitive to changes in network similarity scores. For other subsets of data the results remain similar.

To explore this area further, we created heatmaps for each attack and its transferred attack success rate for each source and target model. Using this method and the results from earlier, we should not see a clear pattern for an individual attack that correlates with the network similarity scores, but we should be able to identify individual models or model architectures that show a correlation. We focused on the highest correlation values above 0.8, but could not identify a clear pattern visible in the heatmaps. This could be due to the small difference in the success rates of the attacks. 
The PGD and FGSM attacks have some of the highest correlation values for AlexNet and also show the widest range of success rate differences in general, but when looking at just the transferred success rates sourced from AlexNet, the success rates are very similar with a delta of 0.07. Thus, the previously suggested correlations can be attributed to the small difference in success rates and network similarity scores.

\subsubsection{Supervised Learning}
The next evalution method we looked at is the \texttt{DecisionTreeRegressor} from the scikit-learn library~\cite{scikit-learn}. The feature matrix $\mathbf{X}$ was created using the similarity scores for the network pairs and the amount of layers for each network. The target vector $\mathbf{y}$ contained the success rates of the transferred adversarial attacks.
The dataset was split into training and evaluation sets to ensure robust model evaluation. The feature matrix $\mathbf{X}$ and the target vector $\mathbf{y}$ were split so that 80\% of the data was used to train the model and the remaining 20\% was reserved for evaluation. This split was performed using the \texttt{train\_test\_split} function from the scikit-learn library~\cite{scikit-learn}, with a random state of 42 to ensure reproducibility.

\begin{figure}
    \centering
    \includegraphics[width=0.99\linewidth]{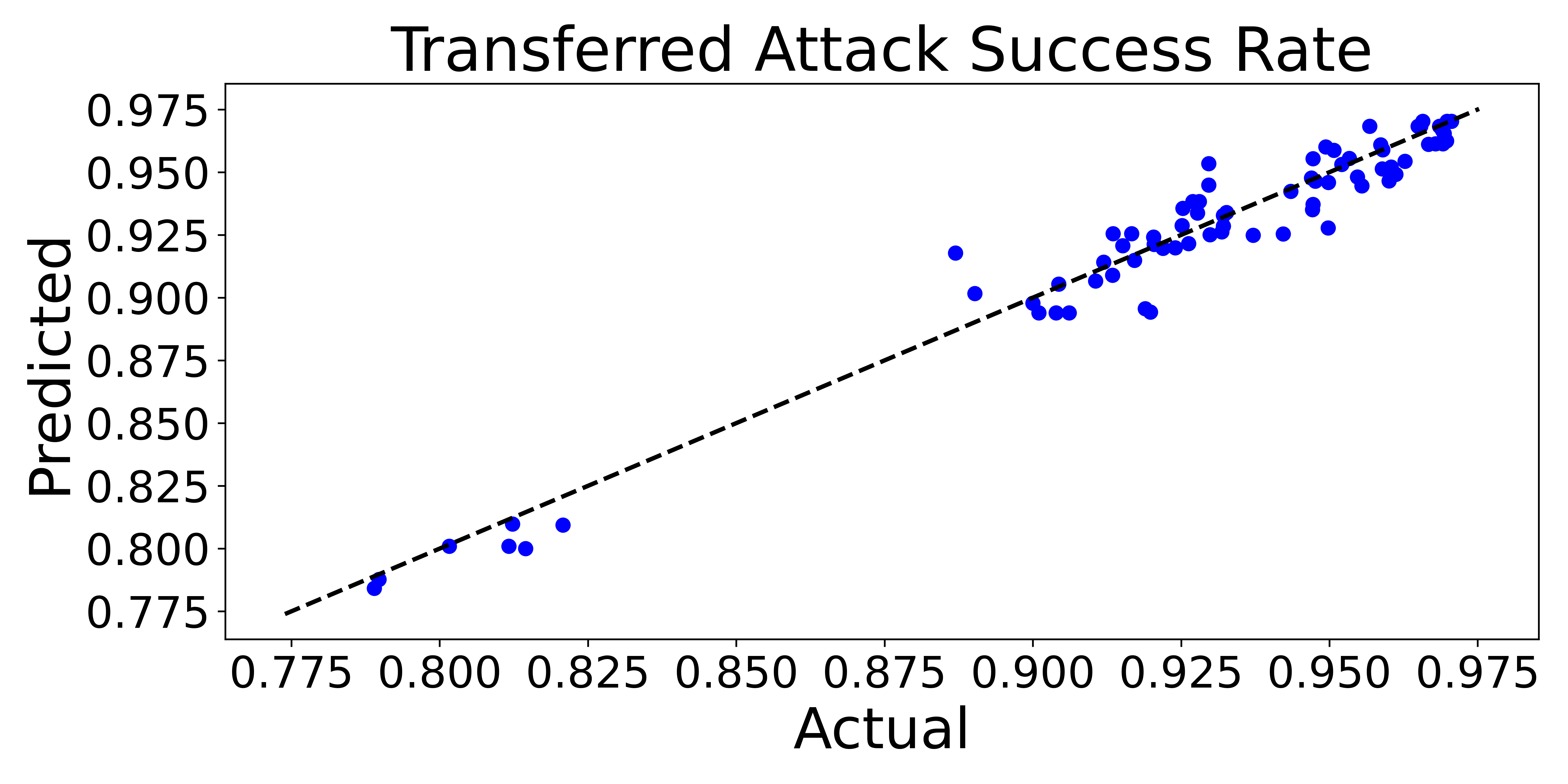}
    \caption{Decision Tree predicted success rates for Black-Box Attacks.} 
    \label{fig:tree_blackbox}
\end{figure}

\begin{figure}
    \centering
    \includegraphics[width=0.99\linewidth]{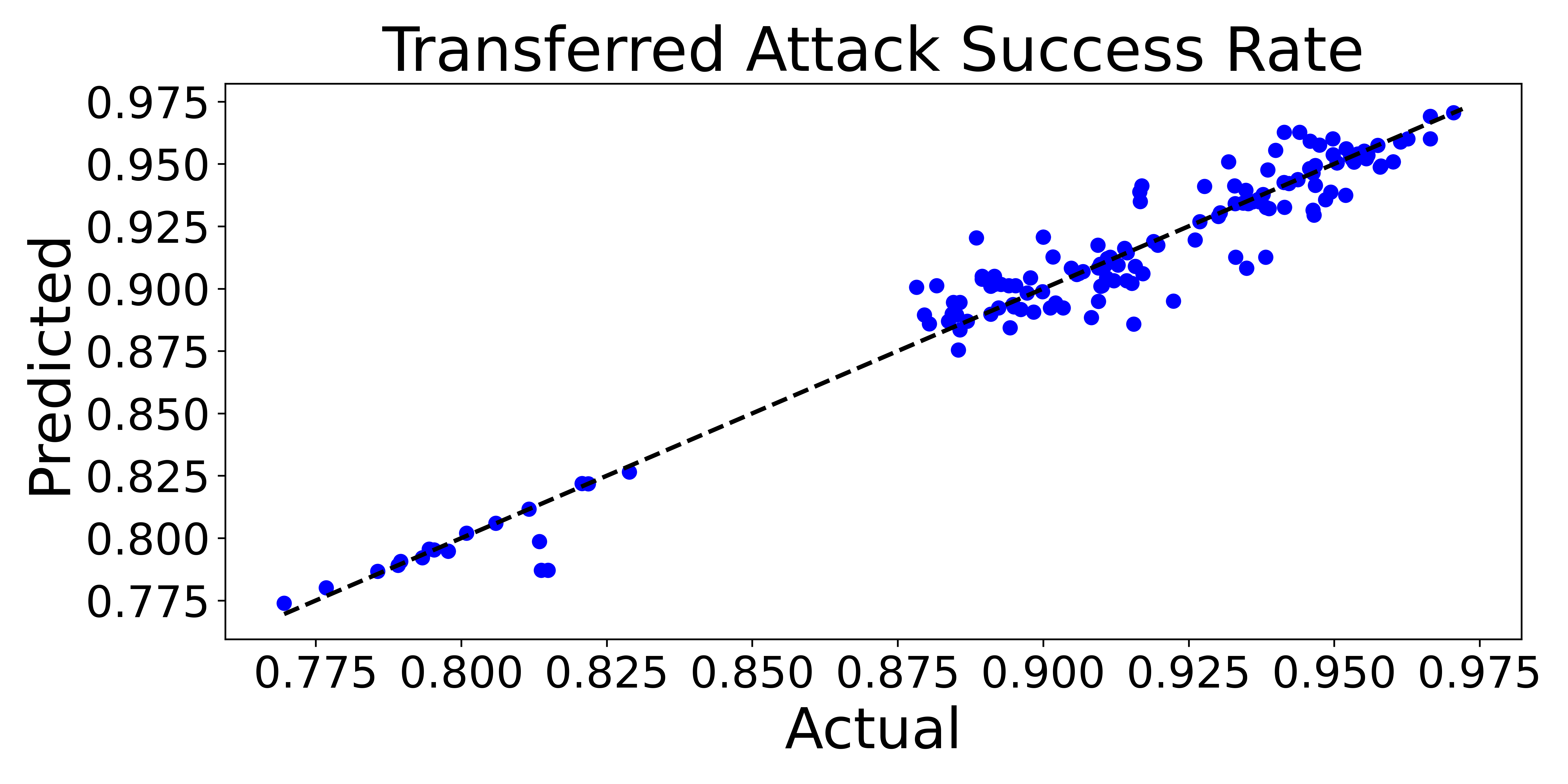}
    \caption{Decision Tree predicted success rates for C\&W Attacks.} 
    \label{fig:tree_carlini}
\end{figure}

The \texttt{DecisionTreeRegressor} shows promising results for two subsets of data. All black-box attacks and the C\&W attacks show an accurate prediction of transferred attack success rate, with a Mean Squared Error (MSE) of close to zero. This indicates that a regression model can predict the success rates of the transferred adversarial attacks from the similarity scores with a high degree of accuracy. You can see the results for the black-box attacks in Figure~\ref{fig:tree_blackbox} and the C\&W attacks in Figure~\ref{fig:tree_carlini}. We can see that all predicted values are close to the actual values, while we see some clustering at the higher and lower success rates. This shows that some of the lower success rates from SqueezeNet and AlexNet were included in the validation set. All other Networks have higher transferred success rates resulting in the bigger cluster at the higher success rates. These results indicate that slight variations in transferred attack success rates can be predicted by the similarity scores for certain attack types. 

\begin{figure}
    \centering
    \includegraphics[width=0.99\linewidth]{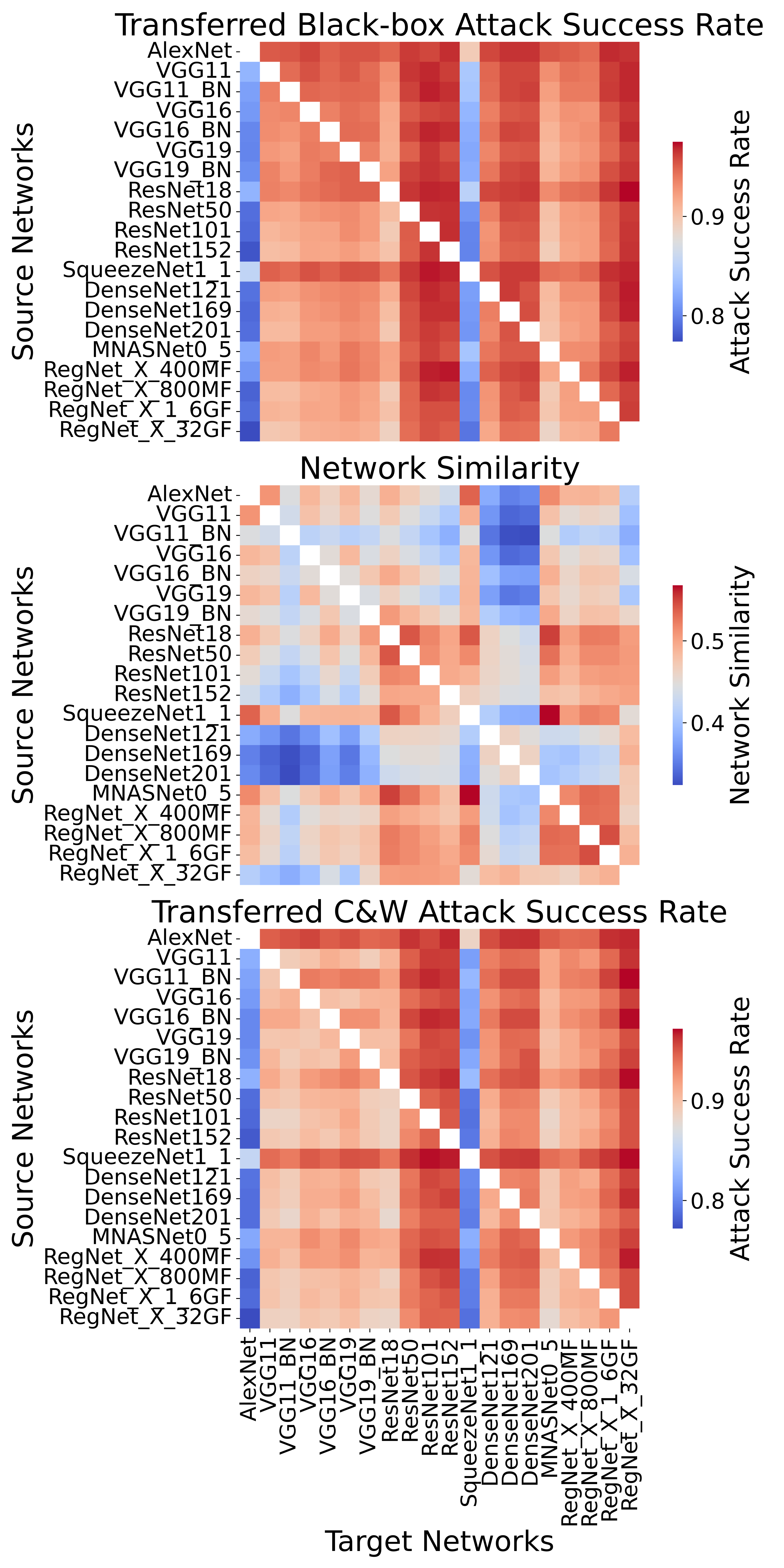}
    \caption{Heatmap for Black-Box Attacks transferred Success Rates, the Network Similarity Scores and C\&W transferred Sucess Rates.}
    \label{fig:heatmap_blackbox}
\end{figure}

Considering these findings we tried to find what causes this correlation. This proves rather difficult because no other subsets of attacks can achieve a similar result. The black-box and C\&W transferred attack success rates can be seen in combination with the network similarity scores in Figure~\ref{fig:heatmap_blackbox}. Looking at these heatmaps and the decission tree results we should be able to identify correlations but upon further investiagation there seems to be no visual tendencies identifiable. The network similarity for SqueezeNet for example shows the highest similarity with MNasNet, but the transferred attack success rate is one of the lowest for this cell.
The same is true for the C\&W attacks. The same example for SqueezeNet applies here as well. 

For the black-box and C\&W attacks the transferred attack success rate for each source network is very similar, with the highest delta being 0.1. But this is true for most other attacks as well, so that can not be a plausible explanation for the high accuracy of prediction.

To confirm this, we tried building a decision tree for all attacks that have a low standard deviation~\ref{tab:adv_attacks_success_rates} (FGSM and PGD were removed), but could not reproduce such high accuracy. For all decision trees we calculated the percentage improvement of the MSE over a mean model, where the prediction for all instances is the mean of the target variable in the training set. In addition we also defined an accuracy metric for the decision tree, we defined a threshhold of 0.01 and counted how many predictions fall within this threshhold and divided by the total number of predictions. The results can be seen in table~\ref{tab:mse_accuracy_comparison_dbs_only}.
These results confirm the high accuracy of the black-box and C\&W attacks, which achieve an accuracy of 73\% and 71\% respectively. The decision tree we build with all attacks, that have a low standard deviation still achieves an MSE improvement of more than 60\% over the mean model, but has an accuracy of only 33\%. Another high percentage improvement is achieved by targeted single-step attacks, which is only FGSM. Targeted FGSM also has a low standard deviation and achieves 27\% accuracy. With these results we feel comfortable in saying that attacks with a standard deviation in transferred attack success of less than 10\% show an MSE improvement of at least 60\%, which means that the predictions are on average much closer to the true values. And the predicted values are within the 1\% threshhold for at least 27\% of the predictions. 
We can also see that black-box and C\&W attacks, when targeted, have an accuracy of 96\% and 91\% with an accuracy gain over the mean model of 73\% and 74\% respectively, further confirming our results.

\begin{table}[h!]
    \centering
    \begin{tabular}{l|c|c|c}
        \hline
        \textbf{Data Subset} & \textbf{Targeted} & \textbf{MSE Improvement} & \textbf{Accuracy} \\
        \hline
        White Box & \xmark & -33.66\% & 3.51\% \\
        \hline
        White Box & \checkmark & -15.38\% & 2.89\% \\
        \hline
        Black Box & \xmark & 95.29\% & 73.68\% \\
        \hline
        Black Box & \checkmark & 99.02\% & 96.05\% \\
        \hline
        Single Step & \xmark & -125.58\% & 3.29\% \\
        \hline
        Single Step & \checkmark & 69.11\% & 27.63\% \\
        \hline
        Multi Step & \xmark & -41.05\% & 6.84\% \\
        \hline
        Multi Step & \checkmark & -21.01\% & 6.25\% \\
        \hline
        Low std attacks & \checkmark & 68.23\% & 35.00\% \\
        \hline
        Low std attacks & \xmark & 60.66\% & 33.42\% \\
        \hline
        All & \checkmark & -1.80\% & 8.08\% \\
        \hline
        All & \xmark & -24.79\% & 3.57\% \\
        \hline
        PGD FGSM & \xmark & -18.65\% & 1.32\% \\
        \hline
        PGD FGSM & \checkmark & -6.31\% & 3.95\% \\
        \hline
        C\&W & \xmark & 94.29\% & 71.05\% \\
        \hline
        C\&W & \checkmark & 97.87\% & 91.45\% \\
        \hline
    \end{tabular}
    \caption{Comparison of MSE Improvement and Accuracy (within 1.0\%).}
    \label{tab:mse_accuracy_comparison_dbs_only}
\end{table}

When looking at the visual representation of the generated decision tree we were not able to identify a clear pattern that could explain the high accuracy. With these results, we can conclude that it is possible to train a model that can predict the success rates of transferred adversarial black-box and C\&W attacks based on the similarity scores. Since black-box attacks have no knowledge of the target network, this could be a valuable tool for network designers to predict the robustness of their networks against adversarial attacks. The limited knowledge of black-box attacks could also be part of the reason for the high accuracy of the decision tree, because these attacks do not have any knowledge of the target network, they seem to be more likely to be influenced by the similarity scores.

\subsubsection{Visual Correlation}

In our efforts to identify potential correlations, we used a comprehensive approach that included several methods. Among these, we prioritised visual analysis of data plots whenever a potential correlation was suggested by other methods. This approach allowed us to assess the validity and significance of these correlations in a more intuitive and direct way. 

Despite these efforts, our careful examination of the visual representations did not reveal any significant findings. Even in cases where other methods indicated strong correlations, visual analysis consistently failed to confirm these results. This lack of visual confirmation highlights the importance of using multiple analytical techniques. It also highlights the potential limitations of relying on visual inspection to confirm complex relationships within data.

%% file: content/conclusion.tex
This study investigated the relationship between network similarity and the success rates of adversarial attacks. Our analysis of the network similarity scores revealed that the neural networks studied exhibited moderate overall similarity, with both the mean and median scores of 0.45 and a low standard deviation of 0.05. This suggests that the networks share a consistent degree of similarity, with little variation in their layer activations, indicating a stable and predictable relationship across different architectures.
Studies could be expanded to include a wider range of network architectures, including those designed for different types of tasks (e.g. natural language processing or reinforcement learning). This would help to determine whether the observed patterns in similarity scores and their contributing factors are consistent across different domains and applications (e.g. images, text or audio).

Further investigation into the influence of network architecture on similarity scores revealed that differences in the number of layers and overall structural complexity play a significant role in determining network similarity. Networks with more complex architectures, such as DenseNet models, consistently had lower similarity scores compared to simpler networks, such as VGG, suggesting that architectural complexity and the presence of unique components, such as DenseBlocks, contribute to reduced similarity. Despite this splitting the data into small (fewer than 200 layers) and big models had no influence on the correlation analysis.

Delving into the similarity of individual layers within these networks. This analysis shows that DataParallel layers have the highest similarity scores, likely due to their consistent position as the first layer in all networks. Dropout, Linear and AdaptiveAvgPool2d layers also show high similarity, which can be attributed to their roles in regularization, classification and feature extraction respectively, and their typical placement towards the end of the networks. Conv2d, ReLU and BatchNorm2d layers also show high similarity due to their fundamental role in CNN architectures. Conversely, specialized and intermediate layers show greater variability and lower similarity scores, reflecting their less consistent role in different networks.

The introduction of the Diagonal Box Similarity (DBS) score provided a refined metric for assessing network similarity by focusing on the most comparable layers between networks. The DBS score, particularly with a box size of 5, showed a wider range of similarity scores compared to the traditional CKA score, providing more nuanced insights into the structural and functional similarities between networks. 
Understanding the architectural features that contribute to higher or lower similarity scores for networks needs to be further investigated. The goal should be to create a more accurate network similarity score that is better able to compare intermediate layers in a network to achieve higher similarity across the same network architectures.

We investigated the success rates of adversarial attacks on different convolutional neural network (CNN) architectures, with a particular focus on understanding their vulnerability to transferred attacks. Our analysis revealed that non-transferred attacks have remarkably consistent success rates across networks, with a mean success rate of approximately 87\% and a standard deviation of 4\%. This consistency suggests that most CNNs are similarly vulnerable to adversarial attacks if the attack is designed specifically for that network, regardless of the network's architecture.
When looking at transferred attacks, we observed more pronounced differences in success rates. In particular the success rates of transferred attacks vary significantly between different networks and attack methods. For example, Projected Gradient Descent (PGD) showed the greatest variation, with success rates ranging from 75\% on AlexNet to 92\% on RegNet\_X\_32GF. In addition, attacks such as FGSM and AutoCG had lower average success rates when transferred, indicating their reduced effectiveness in cross-network scenarios.

Our observations show that when attacks are created on one network, they are more likely to be transferable to other networks if the network is simpler or smaller. When looking at transferability to a network, it is more likely that the attack will be successful if the network is more complex or larger.

The analysis of targeted and non-targeted attacks also revealed interesting trends. Non-targeted transferred attacks generally had lower success rates than their targeted counterparts.

For the main goal of this work, we investigated the relationship between network similarity and the success rates of transferred adversarial attacks using a variety of correlation and analysis methods. We used linear and non-linear correlation techniques, rank correlation measures, and supervised learning models to analyze subsets of data derived from different attack types and network models. Our primary objective was to assess whether the success rates of adversarial attacks can be effectively predicted based on similarity scores.

Our results showed that while there are general trends suggesting a correlation between network similarity and attack success rates, these correlations are not universally strong or consistent across network architectures and attack types.

The DecisionTreeRegressor gave promising results, particularly for black-box and C\&W attacks, achieving an accuracy of over 90\% with an accuracy gain over the mean model of over 73\% for predicting the success rates of these attacks within a 1\% threshold. It seems possible to create a model that can confidently predict the transferability of attacks for a certain network, according to this data.
The inability to reproduce these findings in other subsets indicates that they may be specific to certain types of attacks rather than a universally applicable phenomenon.
This is demonstrated by the observation that subsets characterized by a low standard deviation of transferred adversarial attack success rate achieved an accuracy of approximately 30\%, and the group of white-box attacks achieved approximately 3\%.
Future research can explore other supervised learning models or ensemble methods to improve the robustness and generalizability of the predictions. The goal should be to identify what information from a network and the attacks is required to produce reliable models that can predict the transferability of attacks.

In summary, while this study has provided valuable insights into the correlation between network similarity and adversarial attack success rates, further research is needed to fully understand these relationships and develop more general predictive models.

%% file: main_Arxiv_Version.bbl
%%% -*-BibTeX-*-
%%% Do NOT edit. File created by BibTeX with style
%%% ACM-Reference-Format-Journals [18-Jan-2012].

\begin{thebibliography}{54}

%%% ====================================================================
%%% NOTE TO THE USER: you can override these defaults by providing
%%% customized versions of any of these macros before the \bibliography
%%% command.  Each of them MUST provide its own final punctuation,
%%% except for \shownote{}, \showDOI{}, and \showURL{}.  The latter two
%%% do not use final punctuation, in order to avoid confusing it with
%%% the Web address.
%%%
%%% To suppress output of a particular field, define its macro to expand
%%% to an empty string, or better, \unskip, like this:
%%%
%%% \newcommand{\showDOI}[1]{\unskip}   % LaTeX syntax
%%%
%%% \def \showDOI #1{\unskip}           % plain TeX syntax
%%%
%%% ====================================================================

\ifx \showCODEN    \undefined \def \showCODEN     #1{\unskip}     \fi
\ifx \showDOI      \undefined \def \showDOI       #1{#1}\fi
\ifx \showISBNx    \undefined \def \showISBNx     #1{\unskip}     \fi
\ifx \showISBNxiii \undefined \def \showISBNxiii  #1{\unskip}     \fi
\ifx \showISSN     \undefined \def \showISSN      #1{\unskip}     \fi
\ifx \showLCCN     \undefined \def \showLCCN      #1{\unskip}     \fi
\ifx \shownote     \undefined \def \shownote      #1{#1}          \fi
\ifx \showarticletitle \undefined \def \showarticletitle #1{#1}   \fi
\ifx \showURL      \undefined \def \showURL       {\relax}        \fi
% The following commands are used for tagged output and should be
% invisible to TeX
\providecommand\bibfield[2]{#2}
\providecommand\bibinfo[2]{#2}
\providecommand\natexlab[1]{#1}
\providecommand\showeprint[2][]{arXiv:#2}

\bibitem[Abbasi and Gagn{\'{e}}(2017)]%
        {AbbasiG17}
\bibfield{author}{\bibinfo{person}{Mahdieh Abbasi} {and}
  \bibinfo{person}{Christian Gagn{\'{e}}}.} \bibinfo{year}{2017}\natexlab{}.
\newblock \showarticletitle{Robustness to Adversarial Examples through an
  Ensemble of Specialists}. In \bibinfo{booktitle}{\emph{5th International
  Conference on Learning Representations, {ICLR} 2017, Toulon, France, April
  24-26, 2017, Workshop Track Proceedings}}.
  \bibinfo{publisher}{OpenReview.net}.
\newblock
\urldef\tempurl%
\url{https://openreview.net/forum?id=S1cYxlSFx}
\showURL{%
\tempurl}


\bibitem[Akhtar et~al\mbox{.}(2021)]%
        {meta_adv_attacks}
\bibfield{author}{\bibinfo{person}{Naveed Akhtar}, \bibinfo{person}{Ajmal
  Mian}, \bibinfo{person}{Navid Kardan}, {and} \bibinfo{person}{Mubarak Shah}.}
  \bibinfo{year}{2021}\natexlab{}.
\newblock \bibinfo{title}{Advances in adversarial attacks and defenses in
  computer vision: A survey}.
\newblock
\newblock
\showeprint[arxiv]{2108.00401}~[cs.CV]


\bibitem[{\'A}lvarez et~al\mbox{.}(2023)]%
        {alvarezExploringTransferabilityAdversarial2023}
\bibfield{author}{\bibinfo{person}{Enrique {\'A}lvarez},
  \bibinfo{person}{Rafael {\'A}lvarez}, {and} \bibinfo{person}{Miguel
  Cazorla}.} \bibinfo{year}{2023}\natexlab{}.
\newblock \showarticletitle{Exploring {{Transferability}} on {{Adversarial
  Attacks}}}.
\newblock \bibinfo{journal}{\emph{IEEE Access}}  \bibinfo{volume}{11}
  (\bibinfo{year}{2023}), \bibinfo{pages}{105545--105556}.
\newblock
\showISSN{2169-3536}
\urldef\tempurl%
\url{https://doi.org/10.1109/ACCESS.2023.3319389}
\showDOI{\tempurl}


\bibitem[Andriushchenko et~al\mbox{.}(2020)]%
        {squareattack_bib}
\bibfield{author}{\bibinfo{person}{Maksym Andriushchenko},
  \bibinfo{person}{Francesco Croce}, \bibinfo{person}{Nicolas Flammarion},
  {and} \bibinfo{person}{Matthias Hein}.} \bibinfo{year}{2020}\natexlab{}.
\newblock \bibinfo{title}{Square Attack: a query-efficient black-box
  adversarial attack via random search}.
\newblock
\newblock
\showeprint[arxiv]{1912.00049}~[cs.LG]


\bibitem[Arnab et~al\mbox{.}(2018)]%
        {arnab2018robustness}
\bibfield{author}{\bibinfo{person}{Anurag Arnab}, \bibinfo{person}{Ondrej
  Miksik}, {and} \bibinfo{person}{Philip H.~S. Torr}.}
  \bibinfo{year}{2018}\natexlab{}.
\newblock \bibinfo{title}{On the Robustness of Semantic Segmentation Models to
  Adversarial Attacks}.
\newblock
\newblock
\showeprint[arxiv]{1711.09856}~[cs.CV]


\bibitem[Bhambri et~al\mbox{.}(2020)]%
        {bhambri2020survey}
\bibfield{author}{\bibinfo{person}{Siddhant Bhambri}, \bibinfo{person}{Sumanyu
  Muku}, \bibinfo{person}{Avinash Tulasi}, {and} \bibinfo{person}{Arun~Balaji
  Buduru}.} \bibinfo{year}{2020}\natexlab{}.
\newblock \bibinfo{title}{A Survey of Black-Box Adversarial Attacks on Computer
  Vision Models}.
\newblock
\newblock
\showeprint[arxiv]{1912.01667}~[cs.LG]


\bibitem[Bourzac(2016)]%
        {Bourzac2016}
\bibfield{author}{\bibinfo{person}{Katherine Bourzac}.}
  \bibinfo{year}{2016}\natexlab{}.
\newblock \showarticletitle{Bringing Big Neural Networks to Self-Driving Cars,
  Smartphones, and Drones}.
\newblock \bibinfo{journal}{\emph{IEEE Spectrum}} (\bibinfo{year}{2016}).
\newblock


\bibitem[Brendel et~al\mbox{.}(2018)]%
        {baundary_attack_bib}
\bibfield{author}{\bibinfo{person}{Wieland Brendel}, \bibinfo{person}{Jonas
  Rauber}, {and} \bibinfo{person}{Matthias Bethge}.}
  \bibinfo{year}{2018}\natexlab{}.
\newblock \bibinfo{title}{Decision-Based Adversarial Attacks: Reliable Attacks
  Against Black-Box Machine Learning Models}.
\newblock
\newblock
\showeprint[arxiv]{1712.04248}~[stat.ML]


\bibitem[Bresenham(1965)]%
        {bresenham}
\bibfield{author}{\bibinfo{person}{J.~E. Bresenham}.}
  \bibinfo{year}{1965}\natexlab{}.
\newblock \showarticletitle{Algorithm for computer control of a digital
  plotter}.
\newblock \bibinfo{journal}{\emph{IBM Systems Journal}} \bibinfo{volume}{4},
  \bibinfo{number}{1} (\bibinfo{year}{1965}), \bibinfo{pages}{25--30}.
\newblock
\urldef\tempurl%
\url{https://doi.org/10.1147/sj.41.0025}
\showDOI{\tempurl}


\bibitem[Brown et~al\mbox{.}(2018)]%
        {brownAdversarialPatch2018}
\bibfield{author}{\bibinfo{person}{Tom~B. Brown}, \bibinfo{person}{Dandelion
  Man{\'e}}, \bibinfo{person}{Aurko Roy}, \bibinfo{person}{Mart{\'i}n Abadi},
  {and} \bibinfo{person}{Justin Gilmer}.} \bibinfo{year}{2018}\natexlab{}.
\newblock \bibinfo{title}{Adversarial {{Patch}}}.
\newblock
\newblock
\showeprint[arxiv]{1712.09665}~[cs]


\bibitem[Bunzel et~al\mbox{.}(2024)]%
        {bunzel2024signals}
\bibfield{author}{\bibinfo{person}{Niklas Bunzel},
  \bibinfo{person}{Raphael~Antonius Frick}, \bibinfo{person}{Gerrit Klause},
  \bibinfo{person}{Aino Schwarte}, {and} \bibinfo{person}{Jonas Honermann}.}
  \bibinfo{year}{2024}\natexlab{}.
\newblock \showarticletitle{Signals Are All You Need: Detecting and Mitigating
  Digital and Real-World Adversarial Patches Using Signal-Based Features}. In
  \bibinfo{booktitle}{\emph{Proceedings of the 2nd ACM Workshop on Secure and
  Trustworthy Deep Learning Systems}}. \bibinfo{pages}{24--34}.
\newblock


\bibitem[Bunzel and Graner(2023)]%
        {bunzel2023concise}
\bibfield{author}{\bibinfo{person}{Niklas Bunzel} {and} \bibinfo{person}{Lukas
  Graner}.} \bibinfo{year}{2023}\natexlab{}.
\newblock \showarticletitle{A Concise Analysis of Pasting Attacks and their
  Impact on Image Classification}. In \bibinfo{booktitle}{\emph{2023 53rd
  Annual IEEE/IFIP International Conference on Dependable Systems and Networks
  Workshops (DSN-W)}}. IEEE, \bibinfo{pages}{136--140}.
\newblock


\bibitem[Bunzel and Hamborg(2024)]%
        {bunzel2024adversarial}
\bibfield{author}{\bibinfo{person}{Niklas Bunzel} {and} \bibinfo{person}{Jannis
  Hamborg}.} \bibinfo{year}{2024}\natexlab{}.
\newblock \showarticletitle{Adversarial Patch Detection: Leveraging Depth
  Contrast for Enhanced Threat Visibility}. In \bibinfo{booktitle}{\emph{2024
  54th Annual IEEE/IFIP International Conference on Dependable Systems and
  Networks Workshops (DSN-W)}}. IEEE, \bibinfo{pages}{39--45}.
\newblock


\bibitem[Croce et~al\mbox{.}(2020)]%
        {croce2020robustbench}
\bibfield{author}{\bibinfo{person}{Francesco Croce}, \bibinfo{person}{Maksym
  Andriushchenko}, \bibinfo{person}{Vikash Sehwag}, \bibinfo{person}{Edoardo
  Debenedetti}, \bibinfo{person}{Nicolas Flammarion}, \bibinfo{person}{Mung
  Chiang}, \bibinfo{person}{Prateek Mittal}, {and} \bibinfo{person}{Matthias
  Hein}.} \bibinfo{year}{2020}\natexlab{}.
\newblock \showarticletitle{Robustbench: a standardized adversarial robustness
  benchmark}.
\newblock \bibinfo{journal}{\emph{arXiv preprint arXiv:2010.09670}}
  (\bibinfo{year}{2020}).
\newblock


\bibitem[Csiszárik et~al\mbox{.}(2021)]%
        {csiszárik2021similarity}
\bibfield{author}{\bibinfo{person}{Adrián Csiszárik}, \bibinfo{person}{Péter
  Kőrösi-Szabó}, \bibinfo{person}{Ákos K.~Matszangosz},
  \bibinfo{person}{Gergely Papp}, {and} \bibinfo{person}{Dániel Varga}.}
  \bibinfo{year}{2021}\natexlab{}.
\newblock \bibinfo{title}{Similarity and Matching of Neural Network
  Representations}.
\newblock
\newblock
\showeprint[arxiv]{2110.14633}~[cs.LG]


\bibitem[Das et~al\mbox{.}(2017)]%
        {das2017keeping}
\bibfield{author}{\bibinfo{person}{Nilaksh Das}, \bibinfo{person}{Madhuri
  Shanbhogue}, \bibinfo{person}{Shang-Tse Chen}, \bibinfo{person}{Fred Hohman},
  \bibinfo{person}{Li Chen}, \bibinfo{person}{Michael~E Kounavis}, {and}
  \bibinfo{person}{Duen~Horng Chau}.} \bibinfo{year}{2017}\natexlab{}.
\newblock \showarticletitle{Keeping the bad guys out: Protecting and
  vaccinating deep learning with jpeg compression}.
\newblock \bibinfo{journal}{\emph{arXiv preprint arXiv:1705.02900}}
  (\bibinfo{year}{2017}).
\newblock


\bibitem[Demontis et~al\mbox{.}(2019)]%
        {ambra}
\bibfield{author}{\bibinfo{person}{Ambra Demontis}, \bibinfo{person}{Marco
  Melis}, \bibinfo{person}{Maura Pintor}, \bibinfo{person}{Matthew Jagielski},
  \bibinfo{person}{Battista Biggio}, \bibinfo{person}{Alina Oprea},
  \bibinfo{person}{Cristina Nita-Rotaru}, {and} \bibinfo{person}{Fabio Roli}.}
  \bibinfo{year}{2019}\natexlab{}.
\newblock \showarticletitle{Why do adversarial attacks transfer? explaining
  transferability of evasion and poisoning attacks}. In
  \bibinfo{booktitle}{\emph{Proceedings of the 28th USENIX Conference on
  Security Symposium}} (Santa Clara, CA, USA)
  \emph{(\bibinfo{series}{SEC'19})}. \bibinfo{publisher}{USENIX Association},
  \bibinfo{pages}{321–338}.
\newblock
\showISBNx{9781939133069}


\bibitem[Dong et~al\mbox{.}(2020)]%
        {dong2020benchmarking}
\bibfield{author}{\bibinfo{person}{Yinpeng Dong}, \bibinfo{person}{Qi-An Fu},
  \bibinfo{person}{Xiao Yang}, \bibinfo{person}{Tianyu Pang},
  \bibinfo{person}{Hang Su}, \bibinfo{person}{Zihao Xiao}, {and}
  \bibinfo{person}{Jun Zhu}.} \bibinfo{year}{2020}\natexlab{}.
\newblock \showarticletitle{Benchmarking adversarial robustness on image
  classification}. In \bibinfo{booktitle}{\emph{proceedings of the IEEE/CVF
  conference on computer vision and pattern recognition}}.
  \bibinfo{pages}{321--331}.
\newblock


\bibitem[Goodfellow et~al\mbox{.}(2015)]%
        {fgsm_bib}
\bibfield{author}{\bibinfo{person}{Ian~J. Goodfellow},
  \bibinfo{person}{Jonathon Shlens}, {and} \bibinfo{person}{Christian
  Szegedy}.} \bibinfo{year}{2015}\natexlab{}.
\newblock \bibinfo{title}{Explaining and Harnessing Adversarial Examples}.
\newblock
\newblock
\showeprint[arxiv]{1412.6572}~[stat.ML]


\bibitem[Gupta and Rahtu(2019)]%
        {9010982}
\bibfield{author}{\bibinfo{person}{Puneet Gupta} {and} \bibinfo{person}{Esa
  Rahtu}.} \bibinfo{year}{2019}\natexlab{}.
\newblock \showarticletitle{CIIDefence: Defeating Adversarial Attacks by Fusing
  Class-Specific Image Inpainting and Image Denoising}. In
  \bibinfo{booktitle}{\emph{2019 IEEE/CVF International Conference on Computer
  Vision (ICCV)}}. \bibinfo{pages}{6707--6716}.
\newblock
\urldef\tempurl%
\url{https://doi.org/10.1109/ICCV.2019.00681}
\showDOI{\tempurl}


\bibitem[Haq(2022)]%
        {medical_imaging}
\bibfield{author}{\bibinfo{person}{Imran~Ul Haq}.}
  \bibinfo{year}{2022}\natexlab{}.
\newblock \bibinfo{title}{An overview of deep learning in medical imaging}.
\newblock
\newblock
\showeprint[arxiv]{2202.08546}~[eess.IV]


\bibitem[Hardoon et~al\mbox{.}(2004)]%
        {cca_bib}
\bibfield{author}{\bibinfo{person}{David~R. Hardoon}, \bibinfo{person}{Sandor
  Szedmak}, {and} \bibinfo{person}{John Shawe-Taylor}.}
  \bibinfo{year}{2004}\natexlab{}.
\newblock \showarticletitle{Canonical Correlation Analysis: An Overview with
  Application to Learning Methods}.
\newblock \bibinfo{journal}{\emph{Neural Computation}} \bibinfo{volume}{16},
  \bibinfo{number}{12} (\bibinfo{year}{2004}), \bibinfo{pages}{2639--2664}.
\newblock
\urldef\tempurl%
\url{https://doi.org/10.1162/0899766042321814}
\showDOI{\tempurl}


\bibitem[{K B} and J(2020)]%
        {facial_recognition}
\bibfield{author}{\bibinfo{person}{Pranav {K B}} {and}
  \bibinfo{person}{Manikandan J}.} \bibinfo{year}{2020}\natexlab{}.
\newblock \showarticletitle{Design and Evaluation of a Real-Time Face
  Recognition System using Convolutional Neural Networks}.
\newblock \bibinfo{journal}{\emph{Procedia Computer Science}}
  \bibinfo{volume}{171} (\bibinfo{year}{2020}), \bibinfo{pages}{1651--1659}.
\newblock
\showISSN{1877-0509}
\urldef\tempurl%
\url{https://doi.org/10.1016/j.procs.2020.04.177}
\showDOI{\tempurl}
\newblock
\shownote{Third International Conference on Computing and Network
  Communications (CoCoNet'19)}.


\bibitem[Kornblith et~al\mbox{.}(2019)]%
        {cka_bib}
\bibfield{author}{\bibinfo{person}{Simon Kornblith}, \bibinfo{person}{Mohammad
  Norouzi}, \bibinfo{person}{Honglak Lee}, {and} \bibinfo{person}{Geoffrey
  Hinton}.} \bibinfo{year}{2019}\natexlab{}.
\newblock \bibinfo{title}{Similarity of Neural Network Representations
  Revisited}.
\newblock
\newblock
\showeprint[arxiv]{1905.00414}~[cs.LG]


\bibitem[Lee and Hwang(2022)]%
        {LeeH22}
\bibfield{author}{\bibinfo{person}{Jeonghun Lee} {and}
  \bibinfo{person}{Kwang{-}il Hwang}.} \bibinfo{year}{2022}\natexlab{}.
\newblock \showarticletitle{{YOLO} with adaptive frame control for real-time
  object detection applications}.
\newblock \bibinfo{journal}{\emph{Multim. Tools Appl.}} \bibinfo{volume}{81},
  \bibinfo{number}{25} (\bibinfo{year}{2022}), \bibinfo{pages}{36375--36396}.
\newblock
\urldef\tempurl%
\url{https://doi.org/10.1007/S11042-021-11480-0}
\showDOI{\tempurl}


\bibitem[Li et~al\mbox{.}(2021)]%
        {li2021adversarial}
\bibfield{author}{\bibinfo{person}{Linjie Li}, \bibinfo{person}{Jie Lei},
  \bibinfo{person}{Zhe Gan}, {and} \bibinfo{person}{Jingjing Liu}.}
  \bibinfo{year}{2021}\natexlab{}.
\newblock \showarticletitle{Adversarial vqa: A new benchmark for evaluating the
  robustness of vqa models}. In \bibinfo{booktitle}{\emph{Proceedings of the
  IEEE/CVF International Conference on Computer Vision}}.
  \bibinfo{pages}{2042--2051}.
\newblock


\bibitem[Liu et~al\mbox{.}(2017)]%
        {transferability_0}
\bibfield{author}{\bibinfo{person}{Yanpei Liu}, \bibinfo{person}{Xinyun Chen},
  \bibinfo{person}{Chang Liu}, {and} \bibinfo{person}{Dawn Song}.}
  \bibinfo{year}{2017}\natexlab{}.
\newblock \bibinfo{title}{Delving into {{Transferable Adversarial Examples}}
  and {{Black-box Attacks}}}.
\newblock
\newblock
\showeprint[arxiv]{1611.02770}~[cs]


\bibitem[Madry et~al\mbox{.}(2019)]%
        {pgd_bib}
\bibfield{author}{\bibinfo{person}{Aleksander Madry},
  \bibinfo{person}{Aleksandar Makelov}, \bibinfo{person}{Ludwig Schmidt},
  \bibinfo{person}{Dimitris Tsipras}, {and} \bibinfo{person}{Adrian Vladu}.}
  \bibinfo{year}{2019}\natexlab{}.
\newblock \bibinfo{title}{Towards Deep Learning Models Resistant to Adversarial
  Attacks}.
\newblock
\newblock
\showeprint[arxiv]{1706.06083}~[stat.ML]


\bibitem[Mani et~al\mbox{.}(2019)]%
        {mani2019towards}
\bibfield{author}{\bibinfo{person}{Nag Mani}, \bibinfo{person}{Melody Moh},
  {and} \bibinfo{person}{Teng-Sheng Moh}.} \bibinfo{year}{2019}\natexlab{}.
\newblock \showarticletitle{Towards robust ensemble defense against adversarial
  examples attack}. In \bibinfo{booktitle}{\emph{2019 IEEE Global
  Communications Conference (GLOBECOM)}}. IEEE, \bibinfo{pages}{1--6}.
\newblock


\bibitem[Mi et~al\mbox{.}(2023)]%
        {MI2023114}
\bibfield{author}{\bibinfo{person}{Jian-Xun Mi}, \bibinfo{person}{Xu-Dong
  Wang}, \bibinfo{person}{Li-Fang Zhou}, {and} \bibinfo{person}{Kun Cheng}.}
  \bibinfo{year}{2023}\natexlab{}.
\newblock \showarticletitle{Adversarial examples based on object detection
  tasks: A survey}.
\newblock \bibinfo{journal}{\emph{Neurocomputing}}  \bibinfo{volume}{519}
  (\bibinfo{year}{2023}), \bibinfo{pages}{114--126}.
\newblock
\showISSN{0925-2312}
\urldef\tempurl%
\url{https://doi.org/10.1016/j.neucom.2022.10.046}
\showDOI{\tempurl}


\bibitem[Moosavi-Dezfooli et~al\mbox{.}(2017)]%
        {moosavidezfooli2017universal}
\bibfield{author}{\bibinfo{person}{Seyed-Mohsen Moosavi-Dezfooli},
  \bibinfo{person}{Alhussein Fawzi}, \bibinfo{person}{Omar Fawzi}, {and}
  \bibinfo{person}{Pascal Frossard}.} \bibinfo{year}{2017}\natexlab{}.
\newblock \bibinfo{title}{Universal adversarial perturbations}.
\newblock
\newblock
\showeprint[arxiv]{1610.08401}~[cs.CV]


\bibitem[Nassif et~al\mbox{.}(2019)]%
        {nassif2019speech}
\bibfield{author}{\bibinfo{person}{Ali~Bou Nassif}, \bibinfo{person}{Ismail
  Shahin}, \bibinfo{person}{Imtinan Attili}, \bibinfo{person}{Mohammad Azzeh},
  {and} \bibinfo{person}{Khaled Shaalan}.} \bibinfo{year}{2019}\natexlab{}.
\newblock \showarticletitle{Speech recognition using deep neural networks: A
  systematic review}.
\newblock \bibinfo{journal}{\emph{IEEE access}}  \bibinfo{volume}{7}
  (\bibinfo{year}{2019}), \bibinfo{pages}{19143--19165}.
\newblock


\bibitem[Papernot et~al\mbox{.}(2016a)]%
        {papernotTransferabilityMachineLearning2016}
\bibfield{author}{\bibinfo{person}{Nicolas Papernot}, \bibinfo{person}{Patrick
  McDaniel}, {and} \bibinfo{person}{Ian Goodfellow}.}
  \bibinfo{year}{2016}\natexlab{a}.
\newblock \bibinfo{title}{Transferability in {{Machine Learning}}: From
  {{Phenomena}} to {{Black-Box Attacks}} Using {{Adversarial Samples}}}.
\newblock
\newblock
\showeprint[arxiv]{1605.07277}~[cs]


\bibitem[Papernot et~al\mbox{.}(2017)]%
        {papernotPracticalBlackBoxAttacks2017}
\bibfield{author}{\bibinfo{person}{Nicolas Papernot}, \bibinfo{person}{Patrick
  McDaniel}, \bibinfo{person}{Ian Goodfellow}, \bibinfo{person}{Somesh Jha},
  \bibinfo{person}{Z.~Berkay Celik}, {and} \bibinfo{person}{Ananthram Swami}.}
  \bibinfo{year}{2017}\natexlab{}.
\newblock \bibinfo{title}{Practical {{Black-Box Attacks}} against {{Machine
  Learning}}}.
\newblock
\newblock
\showeprint[arxiv]{1602.02697}~[cs]


\bibitem[Papernot et~al\mbox{.}(2015)]%
        {papernot2015limitations}
\bibfield{author}{\bibinfo{person}{Nicolas Papernot}, \bibinfo{person}{Patrick
  McDaniel}, \bibinfo{person}{Somesh Jha}, \bibinfo{person}{Matt Fredrikson},
  \bibinfo{person}{Z.~Berkay Celik}, {and} \bibinfo{person}{Ananthram Swami}.}
  \bibinfo{year}{2015}\natexlab{}.
\newblock \bibinfo{title}{The Limitations of Deep Learning in Adversarial
  Settings}.
\newblock
\newblock
\showeprint[arxiv]{1511.07528}~[cs.CR]


\bibitem[Papernot et~al\mbox{.}(2016b)]%
        {papernot2016distillation}
\bibfield{author}{\bibinfo{person}{Nicolas Papernot}, \bibinfo{person}{Patrick
  McDaniel}, \bibinfo{person}{Xi Wu}, \bibinfo{person}{Somesh Jha}, {and}
  \bibinfo{person}{Ananthram Swami}.} \bibinfo{year}{2016}\natexlab{b}.
\newblock \bibinfo{title}{Distillation as a Defense to Adversarial
  Perturbations against Deep Neural Networks}.
\newblock
\newblock
\showeprint[arxiv]{1511.04508}~[cs.CR]


\bibitem[Pedregosa et~al\mbox{.}(2011)]%
        {scikit-learn}
\bibfield{author}{\bibinfo{person}{F Pedregosa}, \bibinfo{person}{G Varoquaux},
  \bibinfo{person}{A Gramfort}, \bibinfo{person}{V Michel}, \bibinfo{person}{B
  Thirion}, \bibinfo{person}{O Grisel}, \bibinfo{person}{M Blondel},
  \bibinfo{person}{P Prettenhofer}, \bibinfo{person}{R Weiss},
  \bibinfo{person}{V Dubourg}, \bibinfo{person}{J Vanderplas},
  \bibinfo{person}{A Passos}, \bibinfo{person}{D Cournapeau},
  \bibinfo{person}{M Brucher}, \bibinfo{person}{M Perrot}, {and}
  \bibinfo{person}{E Duchesnay}.} \bibinfo{year}{2011}\natexlab{}.
\newblock \showarticletitle{Scikit-learn: Machine Learning in Python}.
\newblock \bibinfo{journal}{\emph{Journal of Machine Learning Research}}
  \bibinfo{volume}{12} (\bibinfo{year}{2011}), \bibinfo{pages}{2825--2830}.
\newblock


\bibitem[Petrov and Hospedales(2019)]%
        {transferability_1}
\bibfield{author}{\bibinfo{person}{Deyan Petrov} {and}
  \bibinfo{person}{Timothy~M. Hospedales}.} \bibinfo{year}{2019}\natexlab{}.
\newblock \bibinfo{title}{Measuring the {{Transferability}} of {{Adversarial
  Examples}}}.
\newblock
\newblock
\showeprint[arxiv]{1907.06291}~[cs, stat]


\bibitem[Ponn et~al\mbox{.}(2020)]%
        {autonomous_driving}
\bibfield{author}{\bibinfo{person}{Thomas Ponn}, \bibinfo{person}{Thomas
  Kröger}, {and} \bibinfo{person}{Frank Diermeyer}.}
  \bibinfo{year}{2020}\natexlab{}.
\newblock \showarticletitle{Identification and Explanation of Challenging
  Conditions for Camera-Based Object Detection of Automated Vehicles}.
\newblock \bibinfo{journal}{\emph{Sensors}} \bibinfo{volume}{20},
  \bibinfo{number}{13} (\bibinfo{year}{2020}).
\newblock
\showISSN{1424-8220}
\urldef\tempurl%
\url{https://doi.org/10.3390/s20133699}
\showDOI{\tempurl}


\bibitem[Qin et~al\mbox{.}(2023)]%
        {Qin_Xiong_Yi_Hsieh_2023}
\bibfield{author}{\bibinfo{person}{Yunxiao Qin}, \bibinfo{person}{Yuanhao
  Xiong}, \bibinfo{person}{Jinfeng Yi}, {and} \bibinfo{person}{Cho-Jui Hsieh}.}
  \bibinfo{year}{2023}\natexlab{}.
\newblock \showarticletitle{Training Meta-Surrogate Model for Transferable
  Adversarial Attack}.
\newblock \bibinfo{journal}{\emph{Proceedings of the AAAI Conference on
  Artificial Intelligence}} (\bibinfo{year}{2023}).
\newblock
\urldef\tempurl%
\url{https://doi.org/10.1609/aaai.v37i8.26139}
\showDOI{\tempurl}


\bibitem[Raghu et~al\mbox{.}(2017)]%
        {svcca_bib}
\bibfield{author}{\bibinfo{person}{Maithra Raghu}, \bibinfo{person}{Justin
  Gilmer}, \bibinfo{person}{Jason Yosinski}, {and} \bibinfo{person}{Jascha
  Sohl-Dickstein}.} \bibinfo{year}{2017}\natexlab{}.
\newblock \bibinfo{title}{SVCCA: Singular Vector Canonical Correlation Analysis
  for Deep Learning Dynamics and Interpretability}.
\newblock
\newblock
\showeprint[arxiv]{1706.05806}~[stat.ML]


\bibitem[Rashmi and Chaudhry(2024)]%
        {RashmiC24}
\bibfield{author}{\bibinfo{person}{Rashmi} {and} \bibinfo{person}{Rashmi
  Chaudhry}.} \bibinfo{year}{2024}\natexlab{}.
\newblock \showarticletitle{SD-YOLO-AWDNet: {A} hybrid approach for smart
  object detection in challenging weather for self-driving cars}.
\newblock \bibinfo{journal}{\emph{Expert Syst. Appl.}}  \bibinfo{volume}{256}
  (\bibinfo{year}{2024}), \bibinfo{pages}{124942}.
\newblock
\urldef\tempurl%
\url{https://doi.org/10.1016/J.ESWA.2024.124942}
\showDOI{\tempurl}


\bibitem[Roth et~al\mbox{.}(2019)]%
        {roth2019odds}
\bibfield{author}{\bibinfo{person}{Kevin Roth}, \bibinfo{person}{Yannic
  Kilcher}, {and} \bibinfo{person}{Thomas Hofmann}.}
  \bibinfo{year}{2019}\natexlab{}.
\newblock \showarticletitle{The odds are odd: A statistical test for detecting
  adversarial examples}. In \bibinfo{booktitle}{\emph{International Conference
  on Machine Learning}}. PMLR, \bibinfo{pages}{5498--5507}.
\newblock


\bibitem[Sharma et~al\mbox{.}(2018)]%
        {sharma2018analysis}
\bibfield{author}{\bibinfo{person}{Neha Sharma}, \bibinfo{person}{Vibhor Jain},
  {and} \bibinfo{person}{Anju Mishra}.} \bibinfo{year}{2018}\natexlab{}.
\newblock \showarticletitle{An analysis of convolutional neural networks for
  image classification}.
\newblock \bibinfo{journal}{\emph{Procedia computer science}}
  \bibinfo{volume}{132} (\bibinfo{year}{2018}), \bibinfo{pages}{377--384}.
\newblock


\bibitem[Su et~al\mbox{.}(2019)]%
        {transferability_2}
\bibfield{author}{\bibinfo{person}{Dong Su}, \bibinfo{person}{Huan Zhang},
  \bibinfo{person}{Hongge Chen}, \bibinfo{person}{Jinfeng Yi},
  \bibinfo{person}{Pin-Yu Chen}, {and} \bibinfo{person}{Yupeng Gao}.}
  \bibinfo{year}{2019}\natexlab{}.
\newblock \bibinfo{title}{Is {{Robustness}} the {{Cost}} of {{Accuracy}}? --
  {{A Comprehensive Study}} on the {{Robustness}} of 18 {{Deep Image
  Classification Models}}}.
\newblock
\newblock
\showeprint[arxiv]{1808.01688}~[cs]


\bibitem[Suttapak et~al\mbox{.}(2022)]%
        {SUTTAPAK202221}
\bibfield{author}{\bibinfo{person}{Wattanapong Suttapak},
  \bibinfo{person}{Jianfu Zhang}, {and} \bibinfo{person}{Liqing Zhang}.}
  \bibinfo{year}{2022}\natexlab{}.
\newblock \showarticletitle{Diminishing-feature attack: The adversarial
  infiltration on visual tracking}.
\newblock \bibinfo{journal}{\emph{Neurocomputing}}  \bibinfo{volume}{509}
  (\bibinfo{year}{2022}), \bibinfo{pages}{21--33}.
\newblock
\showISSN{0925-2312}
\urldef\tempurl%
\url{https://doi.org/10.1016/j.neucom.2022.08.071}
\showDOI{\tempurl}


\bibitem[Szegedy et~al\mbox{.}(2014)]%
        {szegedyIntriguingPropertiesNeural2014}
\bibfield{author}{\bibinfo{person}{Christian Szegedy},
  \bibinfo{person}{Wojciech Zaremba}, \bibinfo{person}{Ilya Sutskever},
  \bibinfo{person}{Joan Bruna}, \bibinfo{person}{Dumitru Erhan},
  \bibinfo{person}{Ian Goodfellow}, {and} \bibinfo{person}{Rob Fergus}.}
  \bibinfo{year}{2014}\natexlab{}.
\newblock \bibinfo{title}{Intriguing Properties of Neural Networks}.
\newblock
\newblock
\showeprint[arxiv]{1312.6199}~[cs]


\bibitem[Tram{\`e}r et~al\mbox{.}(2017)]%
        {transferability_3}
\bibfield{author}{\bibinfo{person}{Florian Tram{\`e}r},
  \bibinfo{person}{Nicolas Papernot}, \bibinfo{person}{Ian Goodfellow},
  \bibinfo{person}{Dan Boneh}, {and} \bibinfo{person}{Patrick McDaniel}.}
  \bibinfo{year}{2017}\natexlab{}.
\newblock \bibinfo{title}{The {{Space}} of {{Transferable Adversarial
  Examples}}}.
\newblock
\newblock
\showeprint[arxiv]{1704.03453}~[cs, stat]


\bibitem[Virtanen et~al\mbox{.}(2020)]%
        {scypi}
\bibfield{author}{\bibinfo{person}{Pauli Virtanen}, \bibinfo{person}{Ralf
  Gommers}, \bibinfo{person}{Travis~E. Oliphant}, \bibinfo{person}{Matt
  Haberland}, \bibinfo{person}{Tyler Reddy}, \bibinfo{person}{David
  Cournapeau}, \bibinfo{person}{Evgeni Burovski}, \bibinfo{person}{Pearu
  Peterson}, \bibinfo{person}{Warren Weckesser}, \bibinfo{person}{Jonathan
  Bright}, \bibinfo{person}{St{\'e}fan~J. {van der Walt}},
  \bibinfo{person}{Matthew Brett}, \bibinfo{person}{Joshua Wilson},
  \bibinfo{person}{K.~Jarrod Millman}, \bibinfo{person}{Nikolay Mayorov},
  \bibinfo{person}{Andrew R.~J. Nelson}, \bibinfo{person}{Eric Jones},
  \bibinfo{person}{Robert Kern}, \bibinfo{person}{Eric Larson},
  \bibinfo{person}{C~J Carey}, \bibinfo{person}{{\.I}lhan Polat},
  \bibinfo{person}{Yu Feng}, \bibinfo{person}{Eric~W. Moore},
  \bibinfo{person}{Jake {VanderPlas}}, \bibinfo{person}{Denis Laxalde},
  \bibinfo{person}{Josef Perktold}, \bibinfo{person}{Robert Cimrman},
  \bibinfo{person}{Ian Henriksen}, \bibinfo{person}{E.~A. Quintero},
  \bibinfo{person}{Charles~R. Harris}, \bibinfo{person}{Anne~M. Archibald},
  \bibinfo{person}{Ant{\^o}nio~H. Ribeiro}, \bibinfo{person}{Fabian Pedregosa},
  \bibinfo{person}{Paul {van Mulbregt}}, {and} \bibinfo{person}{{SciPy 1.0
  Contributors}}.} \bibinfo{year}{2020}\natexlab{}.
\newblock \showarticletitle{{{SciPy} 1.0: Fundamental Algorithms for Scientific
  Computing in Python}}.
\newblock \bibinfo{journal}{\emph{Nature Methods}}  \bibinfo{volume}{17}
  (\bibinfo{year}{2020}), \bibinfo{pages}{261--272}.
\newblock
\urldef\tempurl%
\url{https://doi.org/10.1038/s41592-019-0686-2}
\showDOI{\tempurl}


\bibitem[Xiao et~al\mbox{.}(2020)]%
        {XiaoZZ20}
\bibfield{author}{\bibinfo{person}{Chang Xiao}, \bibinfo{person}{Peilin Zhong},
  {and} \bibinfo{person}{Changxi Zheng}.} \bibinfo{year}{2020}\natexlab{}.
\newblock \showarticletitle{Enhancing Adversarial Defense by
  k-Winners-Take-All}. In \bibinfo{booktitle}{\emph{8th International
  Conference on Learning Representations, {ICLR} 2020, Addis Ababa, Ethiopia,
  April 26-30, 2020}}. \bibinfo{publisher}{OpenReview.net}.
\newblock
\urldef\tempurl%
\url{https://openreview.net/forum?id=Skgvy64tvr}
\showURL{%
\tempurl}


\bibitem[Xu et~al\mbox{.}(2018)]%
        {Xu0Q18}
\bibfield{author}{\bibinfo{person}{Weilin Xu}, \bibinfo{person}{David Evans},
  {and} \bibinfo{person}{Yanjun Qi}.} \bibinfo{year}{2018}\natexlab{}.
\newblock \showarticletitle{Feature Squeezing: Detecting Adversarial Examples
  in Deep Neural Networks}. In \bibinfo{booktitle}{\emph{25th Annual Network
  and Distributed System Security Symposium, {NDSS} 2018, San Diego,
  California, USA, February 18-21, 2018}}. \bibinfo{publisher}{The Internet
  Society}.
\newblock
\urldef\tempurl%
\url{https://www.ndss-symposium.org/wp-content/uploads/2018/02/ndss2018\_03A-4\_Xu\_paper.pdf}
\showURL{%
\tempurl}


\bibitem[Yin et~al\mbox{.}(2021)]%
        {yin2021adc}
\bibfield{author}{\bibinfo{person}{Mingjun Yin}, \bibinfo{person}{Shasha Li},
  \bibinfo{person}{Chengyu Song}, \bibinfo{person}{M.~Salman Asif},
  \bibinfo{person}{Amit~K. Roy-Chowdhury}, {and} \bibinfo{person}{Srikanth~V.
  Krishnamurthy}.} \bibinfo{year}{2021}\natexlab{}.
\newblock \bibinfo{title}{ADC: Adversarial attacks against object Detection
  that evade Context consistency checks}.
\newblock
\newblock
\showeprint[arxiv]{2110.12321}~[cs.CV]


\bibitem[Zhang et~al\mbox{.}(2019)]%
        {zhang2019adversarial}
\bibfield{author}{\bibinfo{person}{Wei~Emma Zhang}, \bibinfo{person}{Quan~Z.
  Sheng}, \bibinfo{person}{Ahoud Alhazmi}, {and} \bibinfo{person}{Chenliang
  Li}.} \bibinfo{year}{2019}\natexlab{}.
\newblock \bibinfo{title}{Adversarial Attacks on Deep Learning Models in
  Natural Language Processing: A Survey}.
\newblock
\newblock
\showeprint[arxiv]{1901.06796}~[cs.CL]


\bibitem[Żelasko et~al\mbox{.}(2021)]%
        {zelasko2021adversarial}
\bibfield{author}{\bibinfo{person}{Piotr Żelasko}, \bibinfo{person}{Sonal
  Joshi}, \bibinfo{person}{Yiwen Shao}, \bibinfo{person}{Jesus Villalba},
  \bibinfo{person}{Jan Trmal}, \bibinfo{person}{Najim Dehak}, {and}
  \bibinfo{person}{Sanjeev Khudanpur}.} \bibinfo{year}{2021}\natexlab{}.
\newblock \bibinfo{title}{Adversarial Attacks and Defenses for Speech
  Recognition Systems}.
\newblock
\newblock
\showeprint[arxiv]{2103.17122}~[eess.AS]


\end{thebibliography}
